\documentclass{article}
\usepackage[dvipsnames,table,xcdraw]{xcolor}
\usepackage{xcolor}
\usepackage[utf8]{inputenc}
\usepackage{graphicx}
\usepackage{amssymb}
\usepackage{amsmath}
\usepackage{amsfonts}
\usepackage{authblk}
\usepackage{amsthm}
\usepackage[english]{babel}
\usepackage{csquotes}
\usepackage{yhmath}
\usepackage{todonotes}
\usepackage{array}
\usepackage{tocloft}

\usepackage{longtable}
\usepackage{url}
\usepackage[colorlinks=True, urlcolor=black, linkcolor=black, citecolor=black]{hyperref}

\usepackage[margin=1in]{geometry}

\usepackage[backend=biber,sorting=none,style=numeric-comp, bibstyle=numeric]{biblatex}

\title{Assessing the Benefits and Risks of Quantum Computers}

\author[1]{Travis L. Scholten\footnote{Corresponding author; Travis.Scholten@ibm.com}}
\author[2]{Carl J. Williams}
\author[3]{Dustin Moody}
\author[4,5,6]{Michele Mosca}
\author[7]{William ``whurley" Hurley}
\author[8,9]{William J. Zeng}
\author[10]{Matthias Troyer}
\author[1]{Jay M. Gambetta}

\affil[1]{IBM, T.J. Watson Research Center, Yorktown Heights, NY 10598 USA}
\affil[2]{CJW Quantum Consulting LLC, Potomac, MD 20854 USA}
\affil[3]{Computer Security  Division, National Institute of Standards and Technology, Gaithersburg, MD 20854 USA}
\affil[4]{evolutionQ, Waterloo, ON, N2L3G3, Canada}
\affil[5]{Institute for Quantum Computing, University of Waterloo, Waterloo, ON, N2L3G1, Canada}
\affil[6]{Perimeter Institute for Theoretical Physics, Waterloo, Ontario, N2L 2Y5, Canada}
\affil[7]{Strangeworks, Austin, TX 78702 USA}
\affil[8]{Quantonation, 75010 Paris, France}
\affil[9]{Unitary Fund, San Francisco, California 94104, USA}
\affil[10]{Microsoft Corporation, One Microsoft Way, Redmond, WA 98052, USA}

\date{\today}

\begin{document}

\maketitle

\begin{abstract}
Quantum computing is an emerging technology with potentially far-reaching implications for national prosperity and security. Understanding the timeframes over which economic benefits and national security risks (in particular, via cryptanalysis) may manifest themselves is vital for ensuring the prudent development of this technology. To inform security experts and policy decision makers on this matter, we review what is currently known on the potential uses and risks of quantum computers, leveraging currently-available research literature.

The maturity of currently-available quantum computers is not yet at a level such that they can be used in production for large-scale, industrially-relevant problems, and they are not believed to currently pose security risks. We identify 2 large-scale trends -- the development of new approximate methods (variational algorithms, error mitigation, and circuit knitting) and the commercial exploration of business-relevant quantum applications -- which, together, may enable useful and practical quantum computing in the near future.

Crucially, the new approximate methods discussed do not appear likely to change the required resources for cryptanalysis applied to currently-used cryptosystems. From an analysis we perform of the current and known algorithms for cryptanalysis, we find they require circuits of a size exceeding those that can be run by current and near-future quantum computers (and which will require error correction), though we acknowledge improvements in quantum algorithms for these problems are taking place in the literature. In addition, the risk to cybersecurity can be well-managed by the migration to new, quantum-safe cryptographic protocols, which we survey and discuss.

~\\ Given the above, we conclude \textbf{\textit{there is a credible expectation that quantum computers will be capable of performing computations which are economically-impactful before they will be capable of performing ones which are cryptographically-relevant}}.

~\\We hope this work proves informative and useful for those  tasked with making decisions about the use, and future of, quantum computing and quantum-safe technologies.

\end{abstract}

\newpage

\tableofcontents

\newpage

\section*{Paper Motivation and Overview}
\addcontentsline{toc}{section}{Paper Motivation and Overview}
\addtocontents{toc}{%
  \smallskip\protect\parbox[t]{\textwidth}{\textit{Why we wrote this paper, along with a summary of the key themes and topics discussed.}}\par}

\emph{For brevity, this summary omits nuances explicated in the main text.}

Quantum computing poses a known, provable, and substantial risk to the cybersecurity infrastructure that underpins our modern age. Ever since the discovery of Shor's factoring algorithm 30 years ago, the possibility that quantum computers could be used to break commonly-used cryptographic systems has loomed large over the development of the field.

Thus, one of the most pressing questions facing quantum computing is  whether the pursuits taken by industry and academia to build a quantum computer will, at some point, yield the creation of a cryptographically-relevant quantum computer (which can break cryptography) either prior to, or in parallel with, the creation of a quantum computer useful for non-cryptographic applications. This uncertainty is driven by 2 factors.

The first factor is the uncertainty over how the scaling of quantum computers would proceed\footnote{Here, scaling refers to the size (number of gates) of the \textit{quantum circuits} a quantum computer could run. A rough proxy for the size of a quantum circuit is the product of its \textit{width} (the number of qubits it acts on) and its \textit{depth} (the number of primitive layers/timesteps). Note that ``quantum circuits" refers to programs which run on quantum computers.}. Should quantum computers scale rapidly (i.e., faster than the time period over which organizations could transition to post-quantum cryptographic systems), then the risk posed by the development of such computers is substantial. On the other hand, should the scaling proceed more slowly (i.e., organizations have sufficient time to transition their cryptographic infrastructure), then the risk posed, while real and important, is more minimal.

The second factor is that when algorithms for cryptanalysis or useful use cases are studied, the required resources for both kinds of uses appear roughly comparable.  That is, there exist relevant instances of cryptanalysis whose quantum circuits require resources comparable to relevant instances of potentially commercially-useful applications (see Figure \ref{fig:circuit-resource-estimate} in the main text).

Regarding the first factor, many companies have published forward-looking roadmaps describing the capabilities they envision their hardware will have. These roadmaps help make clear the evolution of quantum computing hardware. Many roadmaps envision a key ingredient to scale quantum computers -- namely, quantum error correction --  should be available by the end of this decade. While much research and development needs to take place, the physics and engineering of quantum computers has been de-risked to the point where these roadmaps should be given credence. Given this, we do not focus on this factor in this work. This said, it is important to note that the quantum computers envisioned in the roadmaps would not be useful for cryptanalysis, given what is currently known about the resources required for those types of problems.

Instead, we survey the literature regarding the second factor. The primary uncertainty facing quantum computers over the next few years is whether any applications at all can be realized using current or near-future quantum computers in the absence of (or with minimal) error correction.

Both error-corrected and non-error-corrected quantum computers are noisy, the difference between them being how noisy they are. In general, error-corrected quantum computers (with sufficiently-low physical error rates) can be made \textit{substantially} (exponentially) less noisy than non-error-corrected ones through the use of error-correcting codes. Whether a given use case or application can be realized hinges quite strongly on how much noise can be tolerated in the execution of the requisite circuits before its output is corrupted. The quantum algorithms community has devoted substantial effort to understanding the properties of the circuits needed to realize both useful and cryptographic applications (which we review in Section \ref{sec:benefits}). The prevailing, commonly-held perspective is  ``Realizing quantum applications (of any kind) requires running a single, large-sized quantum circuit, and doing so is only possible with the use an error-correcting code.".

One of the defining features of the algorithms usually considered in those works is the algorithm uses the repeated execution of a single, large-sized circuit. (Here, ``size" means the number of gates in the circuit, a rough proxy for which is the product of the number of qubits the circuit acts on with the number of elementary timesteps of the circuit.) Given this, it is not surprising that error correction is needed: in order to run large-sized circuits accurately, very low noise rates are required.

However, the above analysis also implies that the smaller the size of the circuit required, the more likely the successful execution of that circuit can be realized on current (or near-future) quantum computers and consequently, the realization of any use cases or applications which require those circuits. Thus, the question of whether near-future quantum computers can run useful applications is directly related to the question of whether small-sized circuits can be used to realize them.

From our survey of the research literature, we argue there are 3 major trends which suggest this possibility: the development of new kinds of quantum algorithms which require small-sized or shallow-depth circuits, new techniques for managing the impact of noise when executing quantum circuits, and new methods for decomposing large circuits into smaller ones.  All of these trends utilize the repeated execution of a diverse number of circuits. And all of them are amenable to being used with current and near-future quantum computers. Thus, while realizing quantum error correction continues to be essential and vital for realizing the full potential of quantum computing, it may be possible that near and near-future quantum computers can be put to use to tackle interesting problems, even in the absence of fault-tolerance.

The first method -- new kinds of quantum algorithms -- concerns the development of \textit{variational algorithms} using circuits which can be customized to both the problem at hand and the quantum computer on which the circuit is being run. Importantly, this means these algorithms can be used in a flexible way to scale up the size of the circuits run to the limits of what the hardware could natively support. These algorithms do require the execution of numerous circuits, though. These algorithms can be run on currently-available quantum computers, and should make it easier for quantum computers to run sufficiently-complex circuits as to be useful for scientific or commercial purposes. We discuss variational algorithms in Section \ref{sec:heuristics}.

The second method -- new techniques for managing the impact of noise -- concerns the development of quantum \textit{error mitigation}. These techniques use the repeated execution of an ensemble of circuits, along with classical post-processing, to estimate the output which would be obtained from a noiseless quantum computer. (Phrased another way, they extract more signal from the execution of quantum circuits on noisy hardware.) Error mitigation  can be used with both variational and non-variational algorithms, and extend the reach of the hardware with respect to the size of the circuits it can run.
We discuss quantum error mitigation in Section \ref{sec:error-mitigation}.

The third method -- decomposing large-sized circuits into smaller-sized ones --
is called \textit{circuit knitting}. This is a family of techniques which can be used to reduce the size (or depth) of the circuit by reducing the number of qubits (or timesteps) required, at the expense of needing to run more circuits. The results which would be obtained from a larger circuit are computing using classical post-processing of the results obtained from the set of smaller-sized circuits. Since the circuit size is smaller, they can be run on current and near-future quantum computers, and incur less error, than attempting to run the larger circuit. We discuss circuit knitting in Section \ref{sec:knitting}.

With these techniques, it becomes possible to successfully run circuits on current and near-future quantum computers whose size exceeds that feasible for brute-force classical computation. In addition, it is known that for some problems, small-sized circuits offer provable advantage relative to classical ones. However, both of these results say nothing about whether small-sized circuits are useful for commercial or business applications. Thankfully, a rapidly-expanding commercial ecosystem is exploring the adoption of quantum computing by business. The fact that the commercial ecosystem is engaged in this sort of work should enable the discovery of key business-relevant-problems where a new approach is needed, and the efficacy of quantum is evaluated. Crucially, these explorations can leverage all of the aforementioned advances. Industry is preparing, and many hands make light work of finding commercially-useful applications. We discuss the commercial sector's exploration of quantum computing in Section \ref{sec:commercialqc}.

Thus far, we've noted that variational algorithms, error mitigation, and circuit knitting allow for quantum computing hardware to run circuits whose size (and depth) is larger than what the hardware might otherwise natively be able to support. And we've said the commercial sector can use these methods to explore problems of business relevance. A question arises, though: ``What is the significance of this for cryptanalysis?". We discuss this question in Section \ref{sec:national security}. We find that to date, the literature has not identified a viable path by which these techniques could enable current or near-future quantum computers to attack currently-used cryptosystems. What's more, based on an analysis of the key sizes of currently-used cryptosystems and known algorithms for attacking them (e.g., Shor's), we find that the size of the circuits required to attack them is so large that it is extremely unlikely that such circuits can be run without fault tolerance (see Section \ref{sec:q-algs-pki}).

Further, the remedy to the cybersecurity threat is already known and understood; namely, the need for organizations to transition to quantum-safe cryptographic protocols as soon as possible. We discuss various approaches to doing so in Section \ref{sec:q-safe-approaches}, and outline 
how organizations can begin getting quantum safe in Sections \ref{sec:q-safe-foundations} and \ref{sec:practical-q-safe}.

The themes identified in this survey imply advances in hardware, software, and algorithms are proceeding apace, and it is increasingly likely quantum computers will be useful sooner rather than later. What’s more, given it is highly-probable that the threat to cybersecurity will materialize only after fault-tolerance has been prototyped – and that this development is anticipated to happen in the late 2020s – it is reasonable to surmise that the quantum computers being built for today and the next several years will not beget a cybersecurity risk. Hence, it is probable that scientific or commercial benefits will be attained through the use of quantum computers before cybersecurity risks are realized.

In sum, while the cybersecurity concerns regarding quantum computing are valid and should be taken seriously by agencies and organizations, those concerns should not prevent taking seriously as well the potential end-benefits of using quantum computers for practical problems. Getting both quantum-ready and quantum-safe are crucial activities for agencies and organizations to engage with. For policy makers and regulators, it is crucial to find the “middle way” between the mania of overexuberance in response to quantum hype and the despair of being afraid of cybersecurity threat. Principled and informed regulations and controls can be crafted; such policies need to take a balanced approach to ensuring the end-user- benefits of quantum computers are realized, while ensuring the cybersecurity risks are managed.

\newpage

\section{Introduction}
\addtocontents{toc}{%
  \smallskip\protect\parbox[t]{\textwidth}{\textit{Provides a brief overview of the history of quantum computing, and describes the subsequent sections.}}\par}
  
Although the subject of quantum computation has received much media attention in recent years, the topic is not nearly as new as reports may suggest. One could argue the idea was first introduced by Richard Feynman in a December 1959 speech ``There’s Plenty of Room at the Bottom"\cite{feynman2018there}, where he states:
\begin{quote}
 When we get to the very, very small world---say circuits of seven atoms---we have a lot of new things that would happen that represent completely new opportunities for design. Atoms on a small scale behave like nothing on a large scale, for they satisfy the laws of quantum mechanics. So, as we go down and fiddle around with the atoms down there, we are working with different laws, and we can expect to do different things. We can manufacture in different ways. We can use, not just circuits, but some system involving the quantized energy levels, or the interactions of quantized spins, etc.
 \end{quote}
In this speech, Feynman recognized the quantum-mechanical world provided an opportunity to do things that wouldn't normally be doable at the level of the macroscopic, classical world. (Though it should be noted the use of the word ``circuits" above would \textit{not} have been in reference to quantum circuits, defined previously.) Feynman further elaborated on this idea in his 1981 speech ``Simulating Physics with Computers" \cite{feynman1982simulating}, where he discussed the problem of  using computers to simulate physical systems. His conclusion -- “...and if you want to make a simulation of Nature, you’d better make it quantum mechanical, and by golly it’s a wonderful problem, because it doesn’t look so easy.” -- is a profound one. Even at this early stage of the field of quantum information science, the potential impact of computers operating on quantum-mechanical principles was seen to be just as unprecedented in the history of computation as the introduction of classical computers themselves. Feynman clearly anticipated that quantum computing hardware would be able to efficiently solve or emulate quantum problems, something which appeared to have exponential overhead if done classically. A ``second quantum revolution" -- one in which quantum-mechanical behavior was harnessed and directly put to use -- was beginning to occur, distinct from the advances of the first which had given birth to the semiconductor industry, lasers, and atomic clocks \cite{deutsch2020harnessing,dowling2003quantum}.

In the 1980’s, other individuals began to theoretically explore the concept of, and applications for, a quantum computer. Paul Benioff proposed the idea of a quantum Turing machine -- an idea which ultimately lead to the definition of a universal quantum computer \cite{benioff1980computer}.  This was followed by several theoretically-interesting, yet largely impractical, quantum algorithms including Deutsch’s quantum oracle algorithm \cite{deutsch85}, the Bernstein-Vazirani algorithm \cite{bernstein97}, and Simon’s algorithm \cite{simon1994algorithms}.

Peter Shor’s 1994 discovery of a polynomial time quantum algorithm for factoring integers and computing discrete logarithms \cite{shor1994algorithms} was a breakthrough which stimulated tremendous interest in quantum computing \cite{wiki-QC}. This was the first quantum algorithm to show a super-polynomial speedup over classical algorithms capable of solving the same problem.  Shor’s algorithm was followed by Grover’s unstructured search algorithm  in 1996 \cite{grover1996fast}, followed by a proof of Feynman’s 1981 conjecture showing quantum computers could  simulate quantum systems without the exponential overhead incurred by classical computers \cite{lloyd1996universal}.

While Feynman’s speeches clearly precede Shor’s algorithm, many scientists believe the discovery of Shor’s algorithm stimulated the growing field of quantum information science. Importantly, one of the key results from research following Shor's algorithm was the discovery of \textit{quantum error correction} \cite{shor1995scheme,shor1996quantum}, which established the conditions under which the inherently-fragile information encoded and processed in a quantum computer could be manipulated in a fault-tolerant manner \cite{steane1996simple,calderbank1996good}. (For surveys on the topic of quantum error correction, see \cite{lidar2013quantum,brun2019quantum,RevModPhys.87.307,gaitan2008quantum,Devitt_2013}.) Soon after this discovery was the development of theory of fault-tolerant quantum computation \cite{aharonov1997fault}. Of particular importance was the discovery of the threshold theorem, which established that if the underlying error rates in a quantum computer could be brought low enough, then error-correcting codes could be used to exponentially suppress errors with a polynomial overhead \cite{548464,aharonov1997fault,knill1998resilient,knill1998resilient-2}.

Together, Shor's algorithm and the discovery of quantum error correction further ignited interest in quantum information science. This can clearly be seen by the explosive growth of papers and articles on quantum information and quantum computation beginning around 1994. Crucially, this meteoric growth also encompassed advances in harnessing and controlling quantum-mechanical systems to make them do computation, with the first quantum gate in 1995 \cite{monroe1995demonstration} based on a concept introduced earlier that year \cite{cirac1995quantum}, followed by a first demonstration on Shor's algorithm on a liquid state Nuclear Magnetic Resonance system \cite{vandersypen2001}. Shor’s algorithm, along with the proof of Feynman’s 1981 conjecture and the discovery of quantum error correction, implied that a quantum computer, if successfully built, would have extraordinary impact on society and mankind.

As a result of those early efforts, numerous groups began exploring the possibilities of building quantum hardware based on various underlying qubit modalities (types)\footnote{The qubit (``que-bit") is the fundamental unit of information in a quantum computer, and is the quantum analogue of the classical bit.}$^{,}$\footnote{Modalities currently being investigated include superconducting circuits \cite{Bravyi_2022,Kjaergaard_2020}, trapped ions \cite{monroe2013scaling,HAFFNER_2008,Bruzewicz_2019}, neutral atoms \cite{Henriet2020quantumcomputing, saffman2016quantum}, silicon \cite{Burkard_2023,Chatterjee_2021}, photonics \cite{Flamini_2018,O_Brien_2009,slussarenko2019photonic}, and several others.}.  Simultaneously, but at a slower rate of progress, scientists developed and improved the understanding of quantum complexity classes, and initiated the difficult process of discovering other possible classes of algorithms benefiting from the existence of a quantum computer.   For most of the next two decades, the exploration of quantum hardware, algorithms, and software largely proceeded independently.  It was only in the mid 2010’s that full-stack quantum computing efforts emerged.

Against this backdrop, the spectre of whether, and/or when, a \textit{crytographically-relevant quantum computer} -- one which would threaten currently-used cryptographic systems -- could be built has loomed large. (An equivalent phrase also sometimes used is ``cryptanalytically-relevant quantum computer".) In the early days of quantum computing, sufficiently little was known about the potential benefits of quantum computers (besides their use for simulating quantum-mechanical systems) that identifying whether the economic benefits outweighed the national security risks was difficult. Over the past decade though, a proliferation of work on the applications of quantum computers to problems of commercial and scientific interest has taken place. This work informs the analysis we give here.

The remainder of this paper will begin with a review of our broad understanding of the  potential benefits the use of quantum computers may bring to society.  We describe the rapid advances being made and the best understanding of the required capabilities needed by a quantum computer to bring such benefits in Section \ref{sec:benefits}.  In Section \ref{sec:national security}, we discuss the impact of quantum computing on national security, by reviewing of the evolution of public-key cryptography -- as this appears to be the most pressing risk created by quantum computers -- in Section \ref{sec:pki}, and providing  a careful analysis of the capabilities a quantum computer would need to break current public-key cryptosystems in Section \ref{sec:q-algs-pki}.  In Section \ref{sec:quantum-safe} we discuss the advent of post-quantum cryptography and the steps organizations can take to begin migrating their cryptographic systems to ensure information safety and security as quantum computing systems mature. We conclude in Section \ref{sec:conclusions} with a few comments which we hope will help drive informed discussions amongst security experts and policy makers.

\section{The Political, Cultural, and Scientific Backdrop}
\label{sec:backdrop}

\addtocontents{toc}{%
  \smallskip\protect\parbox[t]{\textwidth}{\textit{Discusses the broad backdrop against which a nascent commercial quantum computing sector has emerged, along with some of the key trends and themes from the past decade.}}\par}
Ever since the early days of quantum computing, the United States government (USG) has provided substantial support and funding to the development of this technology, from a hardware, software, theory/applications, and computational science perspective. The policy decisions taken by the USG have profoundly influenced the trajectory by which quantum computing has developed. We briefly review some of the key policy-related activities before turning to the relatively recent uptake of quantum computing by the private sector.

In 2009, the National Science and Technology Council (NSTC), a Cabinet-level council within the Executive Office of the President responsible for coordinating USG science and technology policy, published the USG's first official policy recommendations for quantum information science \cite{nstc2009}\footnote{It should be mentioned the Department of Defense, in coordination with the National Security Agency, had funded quantum computing research since the mid-1990's and that numerous agencies had previously written reports on both quantum information science and quantum computing.}. This document shows the USG clearly understood quantum technology was going to be a disruptive one, and have an impact on the $21^{\text{st}}$ century. It was understood that sometime in the following decade, industry would begin to invest in this nascent emerging technology. The 2009 NSTC report was followed by another NSTC report in 2016 \cite{nstc2016} identifying quantum information science as a priority for Federal coordination and investment. Finally, in 2018 the NSTC produced a report outlining a set of policy strategies the USG could take to further foster the development of quantum technologies \cite{nstc2018}.  That same year, the National Quantum Initiative Act (NQI Act) was passed  and signed into law. The act provided significant resources to grow the necessary workforce essential to the commercial development and deployment of this technology that was already underway\footnote{In reality, industry and venture capitalists in the United States had already started aggressively moving into the quantum space a few years earlier \cite{jurczak2023investing}.}, and to facilitate coordination amongst the private sector, USG entities, and academia \cite{monroe2019us,raymer2019us,merzbacher2020us}.
Additional history of US policy in quantum computing (and quantum technologies\footnote{``Quantum technologies" usually refers to the trio of quantum computation, communication/networking, and sensing.} more broadly) may be found at \cite{crs-quantum-tech-1} and \cite{crs-quantum-tech-2}.

\begin{figure}[t]
    \centering
    \includegraphics[width=.8\linewidth]{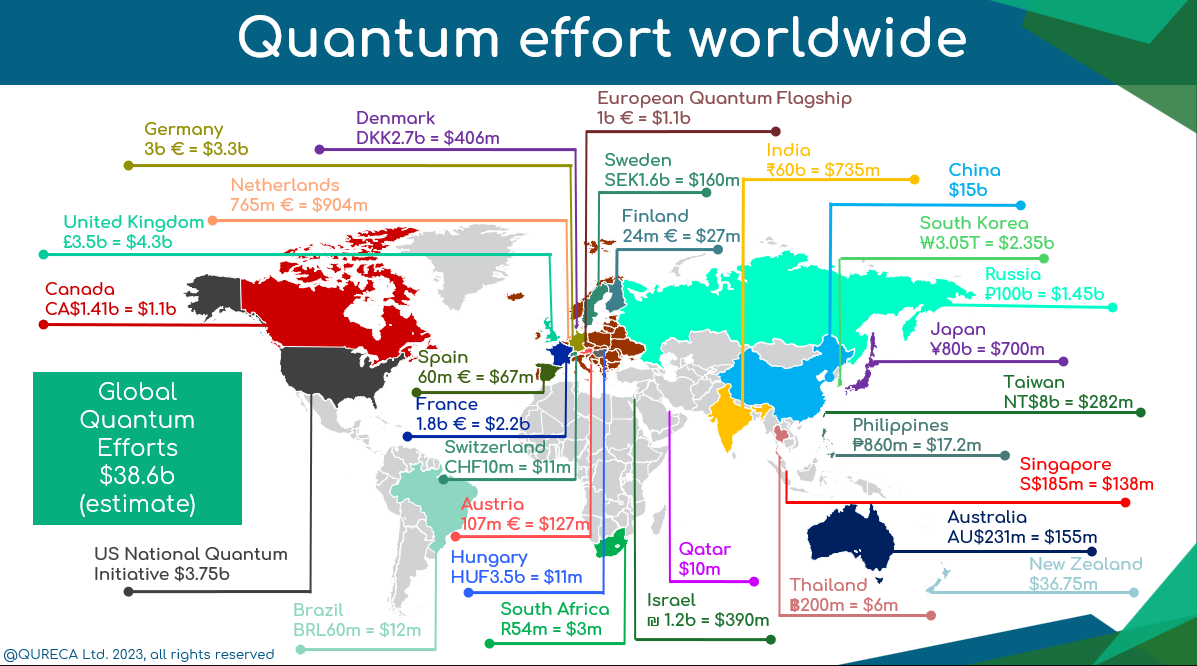}
    \caption{\textbf{Estimated cumulative announced public-sector investments in quantum technologies.} Many governments have announced substantial investments in quantum technologies. The numbers above represent estimates of the total \textit{announced} (not necessarily \textit{deployed}) investment. In addition, these investments span different time horizons, with different start and end dates. Hence, a country-to-country comparison is complicated. Many details and further references are provided in \cite{qureca-investment}, from which this image is taken and used with permission.}
    \label{fig:qureca-investments}
\end{figure}

We note the USG is not alone in developing and establishing a strategy around quantum computing and quantum technologies. Other governments have established whole-of-government plans and strategies, including the European Union (and member states), the United Kingdom, Canada, Australia, New Zealand, Japan, Switzerland, Singapore, Thailand, Qatar, Israel, South Korea, India, South Africa, Brazil, Canada, and China. A summary of these investments may be found at \cite{qureca-investment}. Figure \ref{fig:qureca-investments} presents an estimate of the cumulative announced investments across the globe. Please note this figure shows the total \textit{announced} (not necessarily \textit{deployed}) investment, and that the investments are being pursued over different timescales, depending on the entity involved. Figure \ref{fig:qureca-investments} clearly shows that interest in quantum technologies is strong, and transcends socio-cultural boundaries\footnote{We note in passing that, in addition to the initiatives shown in Figure \ref{fig:qureca-investments}, quantum technologies are also featured prominently in recently-announced international partnerships, being incorporated into ``Pillar 2" of the 2021 AUKUS agreement between Australia, the United Kingdom, and the United States \cite{aukus}, as well as being the subject of several bilateral agreements between the US and various nations \cite{usg-bilateral}. Many other agreements exist as well, either between nation-states or regional ecosystem partners.}.

In the US, the passage of the NQI Act reinforced the importance of quantum technologies for the USG. Its importance to private industry has been made clear by developments starting almost a decade ago. By the mid 2010’s, multiple big-tech and startup companies had created research and development programs to develop quantum computers in the United States \cite{ibm-qx,mohseni2017commercialize}.  Some of these were fully integrated and others were hardware- or software-specific.

The private sector was the source of a disruptive moment in May of 2016, with the launch of the first cloud-based platform for accessing gate-based quantum computers by IBM \cite{ibm-qx}. This event highlighted that quantum computing was moving out of the lab and into the world. This launch was followed a year later by one from Silicon Valley startup Rigetti Computing~\cite{rigetti-launch}. Both launches underscored that venture funded startups, in addition to establish tech companies, were in the quantum computing race. The launches highlighted some of the complexities of deploying quantum computing systems to the cloud in a reliable and robust manner, which has spurred industry to adopt a holistic, systems-engineering-based approach to standing up and deploying quantum computers.

One of the key challenges in making quantum computers accessible (and useful!) to end-users is dealing with the \textit{noise} or \textit{errors}, which degrade the quality of the quantum computation and limit a quantum computer's capabilities. Reducing noise and errors is thus essential. In addition to driving down the rate of errors in the underlying hardware, it is necessary to realize large-scale \textit{error correction} to unlock the full potential of quantum computers from an end-user or application perspective. Error correction can reduce (to arbitrary levels) the effect of noise.  With it, from a large number of noisy physical qubits $n$, a number of logical qubits $k$ which are substantially less noisy\footnote{The exact relationship between $n$ and $k$ depends on a variety of factors, including the error-correcting code used, assumptions about the noise, the physical layout of the qubits, etc. .} are generated\footnote{Roughly speaking, for a code with distance $d_{C}$ (a measure of the number of errors it can correct) and threshold $p_{\mathrm{th}}$ (the maximal error rate below which increasing the code distance $d_{C}$ results in an exponential suppression of the error), if the \emph{bare} or \emph{physical} error rate is $p_{\mathrm{ph}}$, then the \emph{logical} error rate $p_{\mathrm{L}}$ goes as $p_{\mathrm{L}} \sim (p_{\mathrm{ph}}/p_{\mathrm{th}})^{\mathcal{O}(d_{C})}$.}. In addition to generating logical qubits, error-correcting codes define procedures for implementing error-corrected gates on them. 

The ``holy grail" for quantum computing is realizing a fault-tolerant quantum computer. Doing so requires attaining all of the \textit{DiVincenzo criteria} for quantum computing \cite{divincenzo2000physical} -- originally proposed as criteria for physical implementations of quantum computers -- at a \textit{logical} level; see Table \ref{tab:divincenzo}. \textit{\textbf{For a quantum computer to be fault-tolerant, all of these ``logical" DiVincenzo criteria need to be attained simultaneously}}. 

\begin{table}[t]
\centering
\begin{tabular}{|p{8cm}||p{8cm}|}
\hline
   \textit{Physical}  & \textit{Logical} \\ \hline
 A scalable physical system with well-characterized qubits    & A scalable encoding into well-characterized error-correcting codes \\ \hline
 The ability to initialize the system to a simple fixed state & The ability to initialize the system to a simple fixed state and inject magic states \\ \hline
 Long relevant decoherence times & The error-correcting code used suppresses errors far below the physical error rates, and larger codes suppress more errors \\ \hline
 A universal set of quantum gates & A universal set of logical quantum gates \\ \hline
 A qubit-specific measurement capacity & A qubit-specific measurement capacity \\ \hline
\end{tabular} 
\caption{\textbf{The DiVincenzo Criteria}. The physical version of these criteria was introduced in \cite{divincenzo2000physical} to describe the requirements needed to achieve scalable quantum computation. The corresponding version of each criteria applicable to fault-tolerant quantum computation is given.}
\label{tab:divincenzo}
\end{table}

In recent years, a new set of techniques has been developed for dealing with errors in a manner different than error correction. These \textit{error mitigation} techniques were disseminated through a number of pioneering papers \cite{temme2017error, endo2018practical, Li2017_Mitigation} in 2017 and 2018. Error mitigation uses classical post-processing of results obtained from running an ensemble of quantum circuits to estimate what results would have been obtained from a noise-free quantum computer\footnote{The quantum circuit is the fundamental unit of computation for a quantum computer. Circuits act on qubits to encode and process information.}. In addition to the works just noted, subsequent ones from the community have indicated that error mitigation could be a viable path to both unlocking utility from near-term quantum computers, as well as achieving error-corrected quantum computation. We defer an in-depth discussion of error mitigation to the next section. 

As quantum computers were being made accessible to end-users, a publication in 2018 crystallized 3 key needs the quantum computing community (academics and industry) needed to address \cite{Preskill2018quantumcomputingin}:
 \begin{enumerate}
     \item[(1)] Build and deploy quantum computing systems capable of performing tasks which are, for all practical purposes, beyond the capabilities of purely-classical computers.
     \item[(2)] Leverage this hardware to explore problems of scientific and commercial interest.
    \item[(3)] Bridge from near-term towards fully fault-tolerant quantum computing.
 \end{enumerate}

The advent of quantum computing in the cloud, the passage of the NQI Act, and the publication of the above-mentioned work were, collectively, a watershed moment for the quantum computing industry. Big-tech companies continued making substantial investments in their teams. Venture capital took even more of an interest in this technology, and began supporting the nascent quantum industry, with an annual investment exceeding \$2 billion USD in each of 2021 and 2022 \cite{mckinsey-investment,jurczak2023investing} and estimated to reach approximately \$1.5 billion USD in 2023 \cite{jurczak2023investing}. This funding has been deployed abroad across the broad quantum industry, with the largest bets being made on quantum computing hardware and quantum software companies.

Five years later, the quantum computing industry has made substantial progress in realizing the ideas presented in \cite{Preskill2018quantumcomputingin}. Quantum computing systems are being built and deployed that can run quantum circuits whose scale makes exact and direct verification using classical computers difficult. These, and other, systems, have been made accessible to academic and industrial end-users, who have leveraged them to explore possible applications and use cases (see Section \ref{sec:benefits}). Finally, the theory and implementation of quantum error correction has experienced significant progress. We highlight exciting aspects of this progress in the paragraphs below.
~\\~\\
\textit{(1) Build and deploy quantum computing systems capable of performing tasks which are, for all practical purposes, beyond the capabilities of purely-classical computers
}~\\~\\

The quantum systems made available via the cloud in 2016 were not, by any means, capable of performing ``beyond-classical"-type tasks. (And the nature of the kinds of tasks which could be called ``beyond classical" was unclear.) For the systems themselves, the complexity of the circuits which could be run on them -- the circuit width (number of qubits) and circuit size (number of gates) -- was quite small, for at least 3 reasons. First, the systems had a small number of qubits,  limiting the width of the circuits which could be run. Second, the systems had (comparatively) high levels of noise, limiting the size of the circuits which could be run. Third, there was a dearth of available software-defined techniques by which the effect of noise and availability of a limited number of qubits could be managed.

Today, the situation has evolved. Systems are being built with increasing numbers of qubits, and the error rates are going down. Further, software capabilities are being deployed that enable quantum computers with a given qubit count and error rates to execute circuits whose width exceeds the qubit count and whose size is sufficiently high that extracting a signal from the noise is difficult.

A first foray into beyond-classical computation occurred with experiments by Google \cite{arute2019quantum,morvan2023phase} and Xanadu \cite{madsen2022quantum} using superconducting and photonics-based quantum computers, respectively. In addition, Chinese researchers have also demonstrated these experiments on both superconducting \cite{wu2021strong, zhu2022quantum} and photonic \cite{zhong2020quantum,deng2023gaussian} platforms. For a history of these  experiments, see \cite{abughanem2023nisq}. The task performed in these experiments was sampling from the output distribution of random quantum circuits \cite{zlokapa2023boundaries,bouland2019complexity}. (For a review of the task of random circuit sampling and the analysis of its hardness, see \cite{Hangleiter_2023}.) However, while this task is hard for classical computers to do assuming the quantum computer running the circuits is noise-free \cite{Bouland_2018}, the addition of noise changes the difficulty of the task for classical computers \cite{fefferman2023effect}. In addition, advances in classical  techniques have substantially lessened the relative advantages of using a quantum computer for this task \cite{villalonga2020establishing,pan2021simulating,huang2020classical,pednault2019leveraging,oh2023tensor}.

Recently, an additional foray has been pursued with quantum utility experiments by IBM \cite{q-utility-23}. The task performed in this experiment was accurately estimating expectation values (averages) of quantum-mechanical observables for non-trivial circuits based on two-dimensional (2D) spin models. This experiment showed that error mitigation techniques \cite{berg2022probabilistic, kim_scalable_2023} work for quantum computers at the scale of 127 qubits and 2880 gates -- well beyond exact classical simulation of quantum circuits. However, due to the problem considered being relatively simple, it  triggered a number of works exploring classical simulation methods beyond standard approaches \cite{tindall2023efficient, begušić2023fast, kechedzhi2023effective, anand2023classical,rudolph2023classical,liao2023simulation,patra2023efficient}. This stream of results, both from quantum and classical methods, has opened an interesting path where quantum computers could potentially be used to validate and help develop classical simulation methods.

A key difference between these two forays is the way in which noise is handled. In the first foray, noise is dealt with through improvements in hardware. However, the effect of noise is still present in these experiments.  In the second foray, the effect of noise is mitigated, so the experiment produces reliable results which can be trusted. However, due to the potential exponential overhead of the error mitigation methods used, simple comparisons to purely-classical methods do not exist (and remains an open research question).

~\\~\\
\textit{(2) Leverage this hardware to explore problems of scientific and commercial interest.}
~\\~\\
In parallel with these developments, industry has been making quantum computers accessible for academic and commercial users. Companies such as IBM (2016) \cite{ibm-qx}, Rigetti (2017) \cite{rigetti-launch}, Quantinuum (2020) \cite{quantinuum-qcs}, IonQ (2020) \cite{aws-braket}, Xanadu (2020) \cite{xanadu-qcs}, Google (2020) \cite{google-qcs}, Oxford Quantum Circuits (2021) \cite{oqc-qcs}, QuEra (2022) \cite{quera-qcs}, Pasqal \cite{pasqal-qcs} (2022), and Quandela (2022) \cite{quandela-cloud}, have made their systems available via the cloud, and others such as Amazon (2020) \cite{aws-braket}, Strangeworks (2021) \cite{strangeworks}, Microsoft (2021) \cite{msft-azure}, and T-Systems (2023) \cite{t-systems} have enabled a reseller model through their web-based services. As a result, the amount of research on, with, and for quantum computers has exploded \cite{research-growth}. For example, the number of scientific papers which used systems available via IBM's quantum cloud service alone since 2016 is up to around 2,800 at the time of this writing, and the number of circuits executed via that service has exceeded 3 trillion.

Further, industry and academia have defined new methods for evaluating the quality of quantum computers \cite{PhysRevA.100.032328, lubinski2023applicationoriented, chen2023benchmarking, Baldwin2022reexaminingquantum, mezher2022assessing,mckay2023benchmarking} as well as developing frameworks by which performance measures can be consistently evaluated across different quantum computers \cite{amico2023defining}.

We defer to the next Section our discussion of the applications of quantum computers to problems of scientific and commercial interest.
~\\~\\
\textit{(3) Bridge from near-term towards fully fault-tolerant quantum computing.}
~\\~\\
Industry has continued to work on realizing a large-scale, fault-tolerant quantum computation in several ways:
\begin{enumerate}
    \item Pursuing improvements in hardware error rates, to realize higher-quality qubits and operations on them. These improvements have taken place across a wide variety of qubit modalities \cite{Wei_2022,Kandala_2021, Henriet2020quantumcomputing,barnes2022assembly,wintersperger2023neutral,moses2023race,Jia_2023,Fang_2022,tanttu2023stability,kumar2021mitigating,wagner2023benchmarking}. Improving the quality of the qubits and operations is necessary for realizing fault-tolerant quantum computing, as well as unlocking near-term utility.
    \item Demonstrating some of the fundamental techniques required for error correction \cite{gupta2023encoding, ryan2022implementing,PhysRevX.11.041058,sivak2022real,google2023suppressing,chen2022calibrated,sundaresan2022matching,brown2023advances,barber2023realtime,bluvstein2023logical}. These constitute the first steps toward unambiguous demonstration of logical qubits, and as the previously-outlined ``logical" DiVincenzo criteria are implemented, we expect to see very small numbers of logical qubits used to run simple quantum algorithms in a fault-tolerant manner in the near future.
    \item Advancing the theory of quantum error correction to yield insights about new families of quantum error-correcting codes which have more favorable properties than commonly-studied ones, and whose properties also most closely match classical codes\footnote{This will likely be crucial for scaling fault-tolerant systems, as it is known that any planar code satisfies $k/n \leq 1/(cd^2)$ for code distance $d_{C}$ and a constant $c$ \cite{bravyi_2010}. That is, for a fixed number of physical qubits $n$, as the code distance $d_{C}$ is increased -- and hence the code could correct more errors -- the number of logical qubits $k$ decreases. Breaking this bottleneck requires developing new codes.}$^{,}$\footnote{See \cite{qec-zoo} for a detailed list of error-correcting codes.}. For example, recently-discovered ``quantum low-density parity-check" (qLDPC) codes offer more favorable encoding rates (i.e., higher values of the ratio $k/n$) than previously-studied qLDPC codes (such as the surface code \cite{Fowler_2012,Litinski_2019_2}), and the number of errors they can correct is higher \cite{PRXQuantum.2.040101, qLDPC21,higgott2021subsystem,xu2023constantoverhead,zhu2023nonclifford,Lavasani_2019}. In addition, qLDPC codes (and others) could potentially be used to realize a fault-tolerant quantum memory \cite{bravyi2023highthreshold,golowich2023quantum}. These codes are being incorporated into industry roadmaps, and significantly increase the likelihood of the successful introduction of a small-scale error-corrected system by the end of this decade.
    \item Identifying new pathways for bridging the gap from near-term to fault-tolerant quantum computing through the use of error-mitigation for certain sub-routines in error correction (namely, $T$-state distillation) \cite{suzuki2020quantum,piveteau2021}. By combining error mitigation with error correction, it may be possible to more efficiently realize the successful execution of circuits at a scale beyond error mitigation alone \cite{koukoulekidis2023framework,Bultrini_2023,PRXQuantum.3.010345,gonzales2023fault, wahl2023zero}. 
\end{enumerate}

In retrospect, the wide availability of quantum systems in the cloud along with open source software to program them since 2016 constituted a turning point towards creating a quantum computing industry. That activity, the passage of the NQI Act in 2018, and the publication of \cite{Preskill2018quantumcomputingin} helped consolidate this initial spurt. Stimulated by investments at enterprise companies and venture capital firms, this industry has taken up the challenge of making quantum computing useful and practical for the world. We turn next to surveying the research literature about the kinds of problems commercial and scientific groups are exploring, and developing a broad understanding of the resources required to tackle them. As part of this survey, we will also see how the spectre of cryptanalysis has loomed large in the background. We will also look at 4 recent trends in the literature (circuit knitting, variational algorithms, error mitigation, and commercial exploration of quantum computers), and discuss what it might look like to realize near-term value from quantum computers.

\section{Implications of Quantum Computing for National Prosperity}
\label{sec:benefits}
\addtocontents{toc}{%
  \smallskip\protect\parbox[t]{\textwidth}{\textit{Surveys what is known about the uses of quantum computers for scientific or commercial benefit, and discusses 4 major trends in the research literature which may hasten the realization of useful quantum computing.}}\par~\\}

One of the most important advances over the past 5 years has been the development of quantum computing within the private sector. In particular, an end-user base has come into existence, as enterprises seek to leverage quantum computing to transform their businesses \cite{clevelandclinic23,bosch22,quantumnetwork17,creditmutuel22,jpmc22}. In addition, academic groups have made substantial strides in understanding how quantum computers could be used to tackle problems of scientific importance, building on the notion of ``simulating nature" as Feynman articulated almost 40 years ago.
Both of these developments have triggered much work examining the kinds of problems -- commercial or scientific -- where quantum computing could make an impact.

That said, the \textit{exact} timeframe by which quantum computing will deliver \textit{meaningful} impact remains open. One of the ways the research community has tried to address such questions is through \textit{resource estimation}. As the name implies, resource estimation is the task of estimating the required resources needed to realize a given use case or application with a quantum computer. A full survey of resource estimation is beyond the scope of this work\footnote{Though we do provide in Table \ref{tab:resource-estimates} in Appendix  \ref{sec:circuit-resource-table} some resource estimates, and a reference summary of some public resource estimation results is available at~\url{https://metriq.info/progress}}. For our purposes it suffices to note that resource estimation efforts may be bucketed into 3 broad categories:
\begin{itemize}
\item \textit{Logical} resource estimates, which estimate the properties of the circuits needed by the application. Usually these estimates are in terms of the number of qubits required, the number of primitive operations (gates) needed, and/or the total algorithmic runtime (depth). The estimates involve assumptions as to the allowed operations by the computer; namely, whether some operations are considered ``native", or whether they need to be re-written in terms of other operations. That is, these estimates could also be seen as originating at the level of the instruction set architecture (ISA) for the quantum computer, after a given circuit has been compiled to that ISA.
\item \textit{Physical} resource estimates, which estimate the physical properties of a quantum computer needed to realize high-accuracy implementations of the required circuits. These estimates are usually derived by compiling the logical resource estimate to an underlying error-correcting code. The output of these estimations is usually the number of physical qubits required, and the amount of wall-clock time which would be needed to run the circuits. Depending on the assumptions which go into this analysis (the code used, spatial layout of the qubits, classical data transfer constraints, physical error rates, etc.) these estimates vary greatly, even given the same underlying circuit \cite{beverland2022assessing,vandam2023using,9951244,8728077,6657074}.
\item \textit{Practical} resource estimates, which estimate end-user-relevant considerations such as the total energy consumption or cost required to realize a given application \cite{fellousasiani2022resource,parker2023estimating}. These estimates can be derived from either starting with physical resource estimates and making additional assumptions (i.e., how much energy is required to operate a single physical qubit) \cite{parker2023estimating}, or fundamental physics/thermodynamics considerations \cite{Fellous_Asiani_2023,Auff_ves_2022}.
\end{itemize}

Moving from the first to the third type of resource estimate involves making additional assumptions about the operation of the quantum computer. While all 3 categories are useful, logical resource estimates hew most closely to the underlying algorithms (and algorithmic primitives) of quantum computing, and allow us to most clearly and cleanly understand what applications have the greatest potential and ease of being realized. These estimates also show most easily how improvements in the underlying quantum algorithms lessen the resources required. So in this work, of these three, we focus on the first. 

In contrast, physical resource estimates -- being derived from the compilation of circuits onto error-correcting codes -- are highly susceptible to change as the underlying error-correcting code does. At first glance this would be useful: these estimates generally fall over time (i.e., fewer physical qubits and wall-clock time are required) and could suggest  quantum computers can realize applications sooner rather than later (extrapolating from industry-published roadmaps on progress for building systems with increasing numbers of qubits). However, most of the physical resource estimates in the literature today are based on one particular error-correcting code; namely, the surface code \cite{Fowler_2012,Litinski_2019_2}. As noted in the previous section, there is a need to study and develop new error-correcting codes with more favorable properties. Hence, if the underlying error-correcting code which will be used by a future quantum computer is \textit{not} the surface code, then \textit{few} of these existing physical resource estimates would be useful for identifying the crossover point between the resources required to realize a given application and the capabilities of a given quantum computer.

Finally, practical resource estimates are few and far between. And while considerations of the energetic or monetary cost of quantum computation are useful and salient for end-users \cite{fellousasiani2022resource,Fellous_Asiani_2023,Auff_ves_2022,parker2023estimating,meier2023energyconsumption,liu2023potential}, they do not yield any indication or guidance as to what applications are most easily realized using quantum computers.

So, in this work, we focus on logical resource estimates. Estimates of this type usually report the number of qubits required, the circuit depth, and/or the number of \textit{Toffoli} or $T$ gates required.  A variety of resource estimates have been done for specific applications \cite{babbush2021focus,reiher2017elucidating,blunt2022perspective,beverland2022assessing,von2021quantum,steudtner2023fault,chakrabarti2021threshold,gidney2021,cruise2020practical,steudtner2023faulttolerant,webber2021impact,Kim_2022,cortes2023faulttolerant,steudtner2023faulttolerant,rubin2023quantum,rubin2023faulttolerant,Babbush_2023,Goings_2022,watson2023quantum,berry2023quantum,wang2023option}. We forgo a full enumeration of all existing resource estimates, and instead present in Table \ref{tab:resource-estimates} of Appendix \ref{sec:circuit-resource-table} a list of 66 logical based resource estimates for a variety of applications. We have grouped the particular problem studied in each reference into one of five application areas -- physics, simulating nature (chemistry, materials), financial engineering, machine learning, and cryptography -- and have indicated the relevant sub-area the problem falls into. This list is intended to be illustrative, not exhaustive, and we welcome any comments or amendments. 

\begin{figure}
    \centering
    \includegraphics[width=1.05\textwidth]{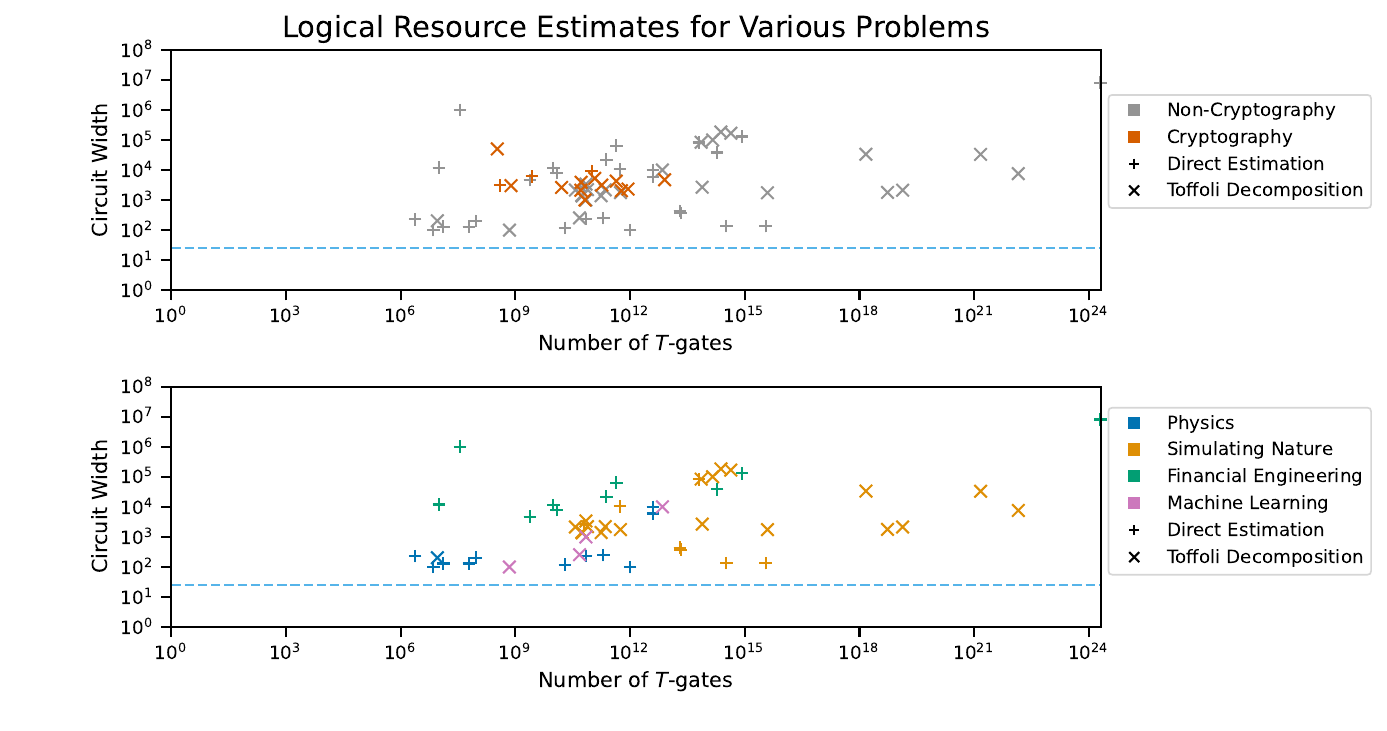}
    \caption{\textbf{Logical resource estimates for various uses of quantum computers.} Plotting the resource estimates of Table \ref{tab:resource-estimates} in Appendix \ref{sec:circuit-resource-table}, where all entries in the table are standardized to $T$ gates, assuming the synthesis of 1 Toffoli requires 7 $T$ gates \cite{gosset2013algorithm}. See main text for a discussion of this choice of synthesis algorithm. Plusses indicate estimates which are done natively with $T$ gates; hatches indicate conversion from Toffoli to $T$. Markers are colored according to the colors of the squares in the legend.  \textbf{Top}: Comparing resources required based on whether the application is relevant for cryptography. The strong overlap between the two colors in the region of $\sim 10^{9}$ to $\sim 10^{12}$ $T$ gates suggests a quantum computer which could run circuits useful for non-cryptographic applications could also \textit{potentially} be cryptographically-relevant. \textbf{Bottom}:  Removing the markers for cryptography and coloring the grey markers in the top figure by application area. Applications in physics (primarily, the dynamics of spin chains) require the fewest resources, both in terms of number of qubits and gates. NOTE: The horizontal line is at a circuit width of 25, a rough and crude limit for circuit widths below which classical simulation is fairly tractable. Simulation remains feasible up to widths of approximately 50, though the compute resources required scale towards those of supercomputers to do so.
    }
    \label{fig:circuit-resource-estimate}
\end{figure}

We plot these resource estimates in Figure \ref{fig:circuit-resource-estimate}. The estimates given in Table \ref{tab:resource-estimates}  involve either Toffoli or $T$ gates. This is because most references report the number of qubits and gates, without focusing on circuit depth. To generate Figure \ref{fig:circuit-resource-estimate}, we explicitly assume the target gate set (allowed/native primitive operations) is the Clifford + $T$ gate set. This gate set is known to be universal for fault-tolerant quantum computation \cite{boykin2000new}, and is one of the most popular gate sets researched in the literature for synthesizing arbitrary quantum operations \cite{niemann2020advanced,jones2013logic,gheorghiu2022t,heyfron2021quantum}. There are many ways to synthesize a Toffoli into a circuit comprised of operations from this gate set \cite{gosset2013algorithm,jones2013logic,heyfron2021quantum}. In the figure, we leverage of the synthesis algorithms of \cite{gosset2013algorithm}, in which each Toffoli is synthesized into a circuit using 7 $T$ gates\footnote{For the expert reader, we acknowledge the algorithm of \cite{PhysRevA.87.022328}, which requires only 4 $T$ gates. However, factors of 2 would not substantially influence the figure, nor the analysis which follows. In addition, the algorithms of \cite{gosset2013algorithm} does not require any additional ancillae, whereas \cite{PhysRevA.87.022328} does.}. We do this acknowledging there are width/gate-count tradeoffs for Toffoli decomposition into $T$ gates \cite{baker2019decomposing,jones2013logic}, as well tradeoffs in terms of optimizations which can be done at the circuit level to minimize either Toffoli or $T$ gate count.

In addition to the above points, other rationales for using the Clifford + $T$ gate set include: (a) realizing high-quality $T$ gates is generally recognized as being a key bottleneck to achieving fault-tolerant quantum computing  \cite{Gidney_2018,Haah2017magicstate,PhysRevA.105.022602,PhysRevLett.118.060501,Litinski_2019},  (b) the classical computational complexity of simulating circuits with $T$ gates generally scales exponentially in their number \cite{bravyi2016improved,Bravyi2019simulationofquantum,Qassim_2021}, and (c) it is known that the output distributions of circuits which include $T$ gates are difficult to model classically \cite{Hinsche_2023}.  These facts, along with the fact the Clifford + $T$ gate set is both universal and popularly-studied, motivates our choice of $T$ gate count as a useful measure of the number of operations required to realize a circuit. 

While the cost of $T$ gates makes them a useful metric, it also encourages algorithmic optimizations to reduce the number of $T$ gates. This can lead to $T$ gates becoming subdominant as in Ref. \cite{von2021quantum}, and then full resource estimations \cite{beverland2022assessing} are needed for more accurate assessments. 

We emphasize that in these resource estimates, it is assumed the algorithm requires a problem-instance-specific, fixed, and single circuit. As we'll discuss below, advances in quantum algorithms, along with new methods for running circuits of large size, actually reduce the size of the circuits required at the cost of needing to run more (and different) circuits. Thus, as advances in quantum algorithms and these methods occur, it may be possible to shift the size of the circuits show in Figure \ref{fig:circuit-resource-estimate} down and to the left.
 
From Figure \ref{fig:circuit-resource-estimate}, four things stand out. The first (via both images) is that there are a fair number of data points on the plot! It is not \textit{a priori} obvious that the research community would have identified any specific problems at all that are amenable to be tackled using quantum computers, much less perform the required resource estimates. And while having more (and more commercially relevant) entries in Table \ref{tab:resource-estimates} would be a boon, it should not be overlooked that there are quite a few ideas in the literature about what one could do with a quantum computer -- although in many of these cases a careful assessment of whether a quantum computer will outperform a classical one \cite{hoefler2023disentangling} still needs to be done. 

The second (via the top image) is that using quantum computers for cryptanalysis requires circuits with properties quite similar to those needed for non-cryptographic uses. That is, the resource estimates do not divide ``cleanly" into cryptographic vs. non-cryptographic applications. Thus, the gap in the resources required between an academic or commercially-relevant application and a cryptographically-relevant one could be argued to be quite small. These analyses could be construed to argue that a quantum computer capable of running circuits useful for some problems of commercial or scientific interest would also be capable of running circuits necessary for cryptanalysis.

The third (via the bottom image) is that of the application buckets we used, applications in physics require circuits with lower complexity than other applications. This suggests that the former might be realized more easily than the latter. However, it is also clear that a quantum computer that can run circuits acting on $\sim100$ to $\sim1000$ qubits can realize a wide variety of applications, assuming it can run on the order of millions to trillions of $T$ gates on those qubits.

And finally, the fourth observation concerns the lack of data points in the region of the figure where the circuit width is below $\sim1000$, and the number of $T$ gates required is $\sim 1K ~ - \sim 10K$. Circuits in this region should, roughly, be achievable by near or medium-term quantum computers. So to date, there is a gap between the capabilities of such quantum computers and what is provably known about what one could do with them when looking at such ``single-circuit" resource estimates.

Based on just the resource estimates in Figure \ref{fig:circuit-resource-estimate}, a question suggests itself as to whether, in the near-term, it might be possible to realize useful applications of quantum computers \textit{without} almost concurrently realizing a cryptographically-relevant threat. A \textit{Nature} news feature from 2019 states that for general applications ``30 years is not an unrealistic timescale” (though others are confident that ``something significant will happen soon.”) \cite{brooks2019beyond}. It is the opinion of the authors that the answer is ``yes", for the reasons discussed in the rest of this section, along with those discussed in Section \ref{sec:q-algs-pki}.

\subsection{Encouraging Trends in the Research Literature}

That the resource estimates in Figure \ref{fig:circuit-resource-estimate} do not divide cleanly into two sets of (namely, cryptographic and non-cryptographic applications) increases the difficulty of determining whether a quantum computer  capable of being used for a variety of non-cryptographic applications is not also concomitantly capable of realizing cryptographic ones. Now, Figure \ref{fig:circuit-resource-estimate} does show some physics-based applications require both comparatively low circuit width and $T$-count relative to cryptanalysis, and some financial engineering applications require lower $T$-count (though with comparable circuit width). So from that perspective, it may be possible to realize useful quantum computing without necessarily realizing a cybersecurity threat.

However, there are additional reasons for optimism. We identify at least 3. First, there have been new advances in quantum algorithms themselves, in which multiple circuits (of lower size) are used. For example, just last year, a new version of Shor's factoring algorithm was proposed \cite{regev2023efficient}. This algorithm requires fewer gates than Shor's algorithm, but requires an additional number of circuits to be run which scales as the square root of the key size.  (We further discuss this algorithm in Section \ref{sec:q-algs-pki}.) Second, new approximate methods for running quantum circuits may shift the results in Figure \ref{fig:circuit-resource-estimate}. These approximate methods are\footnote{Note that approximate methods also exist for fault-tolerant computers \cite{sayginel2023faulttolerant,Suzuki_2020,Wang2022statepreparation,Zhang2022computingground}. }: variational quantum algorithms, error mitigation techniques, and circuit knitting. Third, the commercial sector has begun aggressively investing in the exploration of quantum computers and leveraging access to them. This increases the possibility that heretofore-as-yet-unknown, commercially-relevant applications or use cases may be uncovered.
 
The remainder of this subsection discusses the approximate methods mentioned above, as well as surveys the literature to date on the commercial exploration of quantum computers.

\subsubsection{Variational Algorithms}
\label{sec:heuristics}
\begin{figure}
    \centering
    \includegraphics[width=\textwidth]{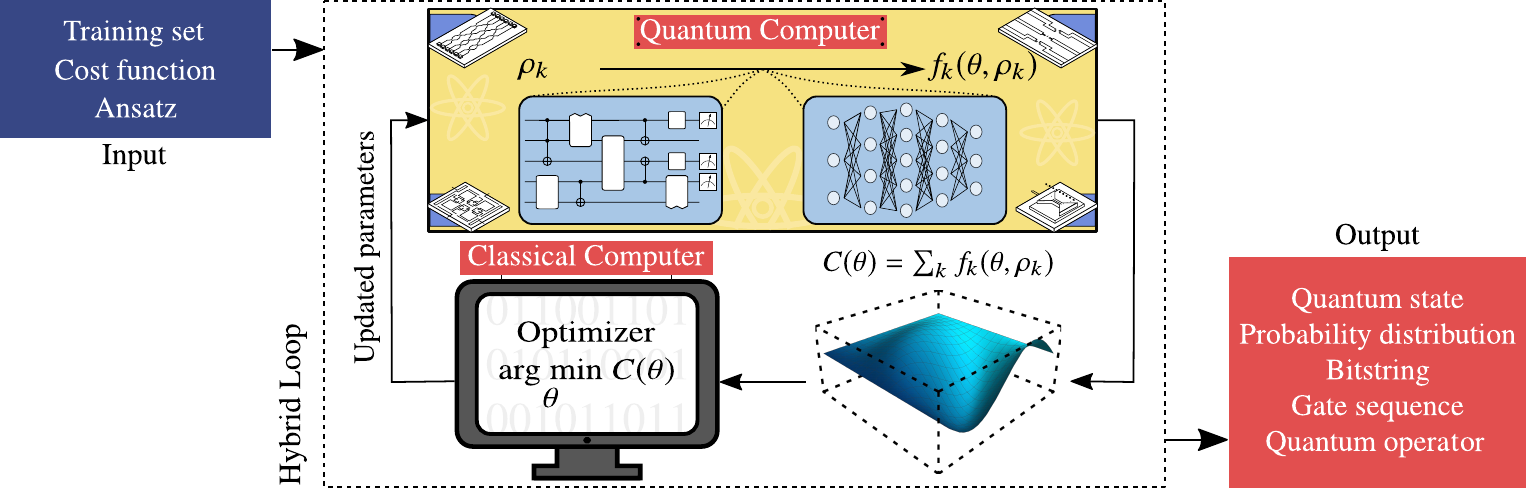}
    \caption{\textbf{Components of a Variational Quantum Algorithm.} A variational algorithm consists of some inputs, a methodology for engaging in an interactive and iterative (``hybrid") quantum-classical compute loop over a timescale exceeding the typical lifetime of the qubits, and produces an output. The purple box shows common inputs to the algorithm. The dashed box shows the loop, which typically features a classical computer adjusting the parameters of a quantum circuit. The red box shows typical outputs once the loop has achieved convergence. NOTE: Image taken from the arXiv version of \cite{cerezo2021variational} with permission.}
    \label{fig:vqa_example}
\end{figure}

The first is the development of \emph{near-term heuristic} or \emph{variational} (hereafter, variational) algorithms specifically designed to be run on near-term quantum computers. At the core of most of these algorithms is a particular integration of classical compute with quantum \cite{cerezo2021variational}\footnote{Note that running algorithms in a fault-tolerant manner \textit{also} requires the integration of purely-classical and quantum compute. The difference being that for variational algorithms, the purely-classical operations can take place on timescales several orders of magnitude larger than those required to realize the error-correction operations necessary for fault-tolerant computers.}. Specifically, these algorithms usually rely on a parameterized trial \emph{ansatz} (i.e., a quantum -- not classical -- guess), denoted here as $|\psi(\boldsymbol{\theta})\rangle$, which is prepared on a quantum computer using a quantum circuit, measured, and which is iteratively improved through adjusting the classical parameters $\boldsymbol{\theta}$ using some kind of optimization method running on a classical computer. These parameters are adjusted based on some cost function $C(\boldsymbol{\theta})$, at which point the circuit is re-run (with the new parameters) until convergence is achieved. After the parameters are optimized, the resulting state might be used to generate a sample (bitstring), estimate an expectation value, etc. (see Figure \ref{fig:vqa_example}). An advantage of these algorithms is that the ansatz can be tailored to both the hardware it runs on and the problem being studied. In addition, the study of variational algorithms generally does not assume the quantum computer running them does so with any degree of error correction.

Although variational algorithms are designed for near-term quantum computers, they could also be run on fault-tolerant ones \cite{blunt2023compilation,sanders2020compilation}. Though in such an implementation, the algorithms would most likely experience a slowdown in runtime due to both the fact that logical operations take longer than their ``bare"/physical counterparts, as well as the fact that the circuits used in these algorithms would need to be compiled to a fault-tolerant gate set.

Examples of variational algorithms include the variational quantum eigensolver (VQE) \cite{peruzzo2014variational,mcclean2016theory,grimsley2019adaptive,baek2022say,tubman2018postponing,baek2022say,PhysRevA.101.010301}, the quantum approximate optimization algorithm (QAOA) \cite{farhi2014quantum,choi2019tutorial,hadfield2019quantum, 8939749, grange2022introduction}, the variational linear systems solver (VQLS) \cite{bravo2019variational,PhysRevA.105.012423,xu2021variational,baskaran2023adapting}, quantum (and quantum-enhanced) models  \cite{schuld2014quest,schuld2017implementing,havlivcek2019supervised,abbas2021power,zoufal2021generative,ghukasyan2023quantumclassical,mangini2023variational}, low depth amplitude estimation~\cite{giurgica2022alg, giurgica2022low, rall2023amplitude, plekhanov2022variational, tanaka2022noisy,Plekhanov2022variationalquantum}, 
and others \cite{bharti2022noisy,chowdhury2020variational,johnson2017qvector,khatri2019quantum,mcardle2019variational,mangini2023variational,Bharti_2022-2,patel2021variational,PhysRevResearch.3.033083,amaro2022filtering}. A thorough discussion of the research to date on variational algorithms is found in  \cite{cerezo2021variational} and \cite{Bharti_2022}. Furthermore, recent work has explored expanding the scope of variational algorithms to include those in which the tunable parameters of the algorithm are offloaded to a classical computer \cite{rebentrost2023}, or where the classical computer uses the outputs from running circuits to develop a surrogate model which can be optimized without the use of parameterized quantum circuits \cite{cerezo2023does}. Such methods may also be rightly considered under the umbrella of variational algorithms.

Fault-tolerant implementations of quantum algorithms under reasonably well-behaved noise models often come with performance bounds which can be proven even in the absence of hardware on which to run them. Similarly, performance bounds for variational algorithms have been devised, though these bounds tend to depend on more contexts/assumptions and be more narrowly applicable than those for fault-tolerant ones \cite{benchasattabuse2023lower,quiroz2021quantifying,Hadfield_2022,Caro_2022,Caro_2021,Thanasilp_2023,gentinetta2022complexity,Wurtz_2021,Basso_2022,ragone2023unified,Liu_2023,letcher2023tight,quiroz2021quantifying,patel2021variational}. In addition, for some variational algorithms and some problems, work does exist analyzing regimes where superior performance may be observed compared to classical methods \cite{farhi2016quantum,hangleiter2022computational,wu2023variational,hibat2023framework,dalzell2020many,kuhn2019accuracy,lee2023evaluating,shaydulin2023evidence,lykov2023sampling,sweke2023potential,hibatallah2023framework,Slattery_2023,K_hn_2019,basso-2022,giurgica2022alg, rall2023amplitude}, or providing insights into the contexts in which variational algorithms offer a useful benefit or advantage \cite{sweke2023potential,Schuld_2022,rudolph2023trainability,Thanasilp_2023,ragone2023unified}. Of course, there are many open questions as to the contexts (specific problem, algorithm, data set used, etc.) in which such variational algorithms could be efficiently simulated or otherwise emulated through purely-classical means \cite{cerezo2023does,xiong2023fundamental,thanasilp2022exponential,goh2023liealgebraic}.

The performance of variational algorithms is often assessed \textit{empirically} through experimentation on hardware. Considered one way, this is an apparent weakness: how can one know whether a given variational algorithm is going to do well for a given problem instance and hardware? The only answer seems to be ``try it and see". Considered another way, this is a strength: these algorithms are amenable to being run on currently-available hardware. Furthermore, the research community has identified classes of shallow-depth circuits which are hard to simulate classically \cite{Bravyi_2018,shepherd2009temporally}, and short-depth circuits have been shown to be applicable to physical experiments \cite{kandala2019error}. However, issues such as barren plateaus can make such circuits difficult to use in a variational algorithm \cite{ragone2023unified,diaz2023showcasing,Arrasmith_2022,CerveroMartin2023barrenplateausin}. It is an open question as to what sort(s) of circuits should be used: too much structure in the circuit makes it easier to simulate classically, and if they are fairly unstructured (i.e., close to random), then finding optimal parameters is difficult \cite{Holmes_2022, cerezo2023does}.

In the absence of well-characterized noise models describing how the hardware behaves, predicting analytically and in advance the behavior of algorithms is hard. So being able to run these algorithms and explore in practice how they behave can help one get a handle on them. This is similar to some advances in purely-classical computing, in which hardware helped improve algorithms, and also led to the discovery of new ones\footnote{For example, the linear-scaling behavior of the simplex algorithm in practice (while the worst-case scaling behavior was exponential) led researchers to develop the notion of \emph{smoothed analysis} as a measure of computational complexity (and under which, the simplex algorithm has polynomial complexity) \cite{spielman2004smoothed}. Or, numerical methods for calculating properties of spin chains led to the development of the renormalization group method \cite{RevModPhys.47.773}.}.

Variational algorithms require hardware which is ``good enough" to run them with high accuracy. Of course, better (i.e., less noisy) hardware makes it easier to attain better performance with such algorithms, and in addition, variational algorithms do have some resilience to noise \cite{fontana2021evaluating,sharma2020noise}. One way to estimate a low enough error rate such that the algorithm can be successfully run is to assume that at each time step of the algorithm, each qubit has an independent and identical probability of suffering an error. In turn, this means the number of ``error opportunities" of an algorithm  requiring a given number of qubits $q$ and having depth $d$ is $q*d$, which can be used as a rough proxy for the size of the circuit\footnote{The subtlety here is that the total number of gates which could be fit in a circuit is upper bounded by this quantity (saturated when the circuit consists of $d$ layers of single-qubit gates on all $q$ qubits in each layer), but depending on how the gates are structured in the algorithm, the actual number could be quite a bit less.}. The inverse $(q*d)^{-1}$ is a loose upper bound on the acceptable error rate successfully run the algorithm. Estimates with this flavor (primarily, inverse in the circuit size) have been used for estimating the success probability of variational algorithms \cite{brandhofer2021special,brandhofer2021arsonisq,moll2018quantum,PRXQuantum.4.010309} (In Section \ref{sec:q-algs-pki}, we adopt a similar approach for estimating the number of  error opportunities of quantum algorithms for cryptanalysis, but at the logical level). What's more, for some classical simulation methods, the error rate of the hardware influences the complexity of doing so \cite{shao2023simulating,fontana2023classical,stilck2021limitations}.
Both of these analyses point to the importance of improving error rates of hardware so that variational algorithms can be successfully run at qubit counts and circuit depth which are of meaningful size to end-users, and also in regimes where classical simulation methods may struggle.

In sum, the line of research in variational algorithms has generated insights about the kinds of circuits a near-term quantum computer could run to explore problems in physics, simulating nature, financial engineering, optimization, and machine learning. However, the success of executing these circuits is hindered by the presence of noise. 

\subsubsection{Error Mitigation}
\label{sec:error-mitigation}
\begin{figure}
    \centering
    \includegraphics[width=.67\textwidth]{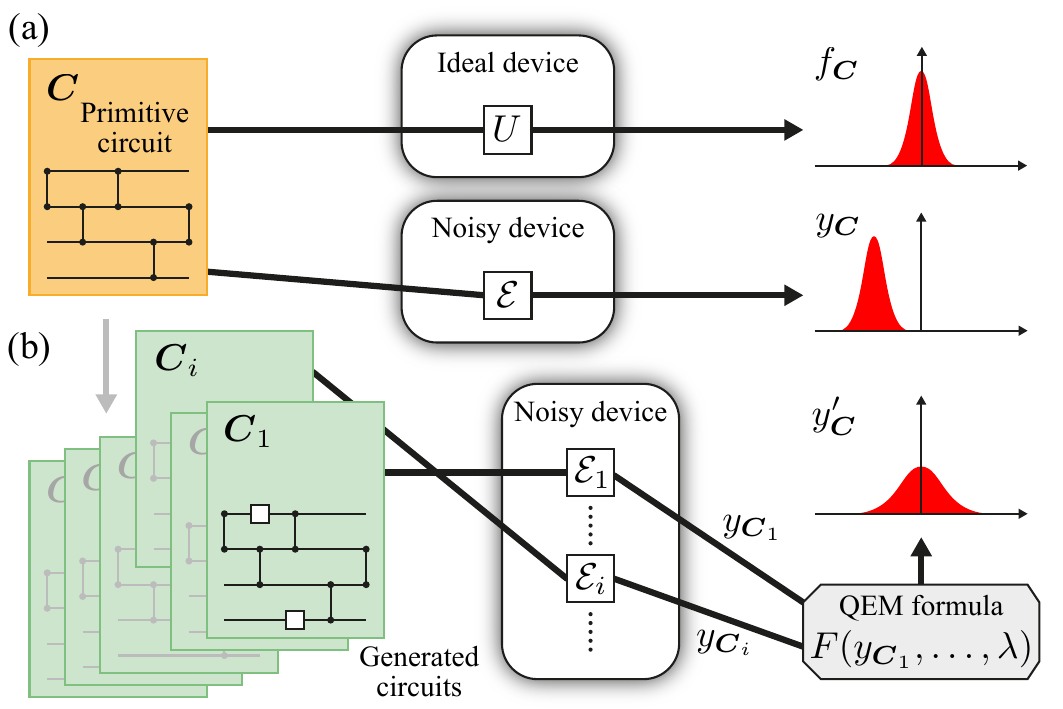}
    \caption{\textbf{Quantum Error Mitigation Overview.}  Quantum error mitigation replaces a single circuit $C$ with an ensemble $\{C_{1}, \cdots , C_{N}\}$, and using a classical post-processing function $F(y_{C_{1}}, \cdots, \lambda)$ on the results of those circuits to estimate the result of what a noise-free quantum computer would output. Here, $f_\mathbf{C}$ is the output of an ideal quantum computer, and $y'_{\mathbf{C}}$ is the output of the error mitigation technique. Some techniques enable unbiased estimates, but at the expense of an increased variance in the estimator. Note: image taken from \cite[Figure 1]{qin2023error} and is used in accordance with the Creative Commons Attribution 4.0 International License \cite{cc-4} under which it was published.}
    \label{fig:qem}
\end{figure}

Thankfully, researchers have also explored techniques to reduce
the effect of noise, without having to resort to improving the error rates of the underlying hardware and the gates they perform. This second line of work has resulted in techniques are called \emph{error mitigation}. At its heart, an error mitigation technique works by replacing 1 instance of a circuit an
 end-user would like to run with an \emph{ensemble} of $N$ circuits, where $N$ is set by the technique itself and the error rates of the hardware. The output from running each of those $N$ circuits are then post-processed (in a manner defined by the technique) to yield an estimate of what an ideal quantum computer would have output when running the original circuit; see Figure \ref{fig:qem}. This makes error mitigation a powerful tool for getting more performance or accuracy from a near-term quantum computer \cite{kandala2019error,russo2022testing,tazhigulov2022simulating,larose2022error,shehab2019toward, mari2021extending, pelofske2023increasing}.

A variety of error mitigation techniques have been developed. Some are used to mitigate the erroneous readout of the state of a qubit (``measurement error mitigation") \cite{nation2021scalable,van2022model,bravyi2021mitigating}. Others used to extrapolate to a zero-noise limit  (``zero-noise extrapolation") \cite{temme2017error,giurgica2020digital}. Still others are used to cancel the effect of the noise (``probabilistic error cancellation") \cite{mari2021extending,temme2017error,berg2022probabilistic,endo2018practical}. Additional techniques exist as well \cite{zhao2023grouptheoretic,liao2023machine,hagge2023error,siegel2023algorithmic,uvarov2023mitigating,jnane2023quantum,obrien2022purificationbased,Gonzales_2023,Bultrini_2023-2}; note more exist than are cited here. With such techniques, end-users of quantum computers can achieve more accuracy in the results they obtain from a quantum computer. However, error mitigation comes with unavoidable tradeoffs \cite{takagi2022fundamental,takagi2022universal,cai2023practical}. Two tradeoffs are the overhead $N$ (usually, exponential in the circuit size) and the variance of the resulting estimate (usually, increased). A full discussion of error mitigation is beyond the scope of this work; see \cite{cai2022quantum,endo2021hybrid,qin2022overview,wang2021error} for more details.

Error mitigation protocols have been used in several recent demonstrations in conjunction with running circuits acting on approximately 100 qubits \cite{chowdhury2023enhancing,shtanko2023uncovering,PhysRevResearch.5.013183, farrell2023scalable}. These demonstrations showcase the usefulness of error mitigation techniques, and highlight how they can be successfully applied to current quantum computers.

With variational algorithms and error mitigation, the underlying hardware on which the algorithms is run has less stringent requirements placed on it to ensure accurate results are obtained. This is because error mitigation allows for the noise to be mitigated, and variational algorithms can be noise resilient. As such, these approaches should allow a quantum computer to run circuits with a higher gate count, thereby advancing along the $x$-axis of Figure \ref{fig:circuit-resource-estimate}.

But what of the $y$-axis? Must we wait until systems with large numbers of sufficient-quality qubits are available before running circuits acting on large numbers of qubits becomes feasible? The answer is ``no", due to the development of certain \textit{circuit knitting} protocols. These protocols, discussed in the next sub-section, provide a way to decompose high-width circuits into numerous smaller-width ones, thereby emulating what a larger-scale quantum computer could do. In addition, circuit knitting can be used to advance along the $x$-axis of the figure, as we will discuss below.

\subsubsection{Circuit Knitting}
\label{sec:knitting}

Circuit knitting refers to protocols whereby a quantum computational problem is broken down into multiple quantum circuits whose outputs are then post-processed (``knit") asynchronously using classical computing. Circuit knitting protocols can be used to extend the circuit width or depth, depending on the protocol. Here, we will discuss on 4 examples of knitting protocols: decomposition protocols, forging protocols, embedding protocols, and time evolution protocols. 

Decomposition protocols are used to decompose a high-width (large-qubit-count) circuit into $N$ smaller-width ones, where $N$ depends on the method and how it is applied. In general, $N$ scales exponentially with the number of partitions used in the decomposition (in a manner to be made more precise below), and this scaling appears to be a general property of decomposition methods, regardless of the specific method used \cite{marshall2023qubit}. In addition, decomposition protocols can be used to emulate circuits of a larger depth through the use of shallower ones \cite{Perez_Salinas_2023}.

Forging protocols leverage the fact that entangled quantum states can be expressed as a linear combination of non-entangled ones (via the Schmidt decomposition), which allow for the calculation of expectation values with respect to entangled states of $2N$ qubits to be re-written as a sum of expectation values of $N$ non-entangled states \cite{Eddins_2022,huembeli2022entanglement}. These methods would be most useful in the context of problems related to simulating the behavior of physical systems (e.g., molecules). The overhead of forging methods depends on the amount of entanglement in the state (as measured by the number of Schmidt coefficients). In the worst case, this overhead is $2^{N}$, but in general, a smaller number of coefficients may be used, substantially lessening the overhead.

Embedding protocols (mostly used in quantum chemistry problems) isolate the most quantum-mechanically difficult part of a simulation, and address that part using a quantum computer \cite{rossmannek2021quantum,rossmannek2023quantum}. (Note that embeddings are used in purely-classical quantum chemistry as well \cite{sun2016quantum}; the novelty of the aforementioned works is the use of a quantum computer as part of the method.) The part of the simulation which is more tractable classically is solved using purely-classical compute, and the two approaches are yoked together to yield a self-consistent and convergent solution to the problem.

Time evolution protocols are used to simulate the time dynamics (evolution) of quantum-mechanical systems. One protocol, introduced in \cite{zhuk2023trotter}, uses multi-product formulas to emulate Trotterized time evolution over a fixed interval using time evolution over smaller intervals. An advantage of this protocol is that the number of circuits required scales linearly with the order of the Trotter evolution. In addition, the approximation error of this method is quadratically improved relative to standard Trotter time evolution approaches, even over arbitrary time intervals.

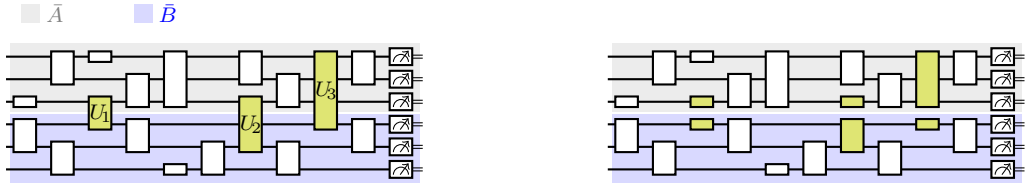
\begin{figure}
    \centering
    \begin{tikzpicture}[thick, scale=1]
    \def\xs{0.07}
    \def \b{0.3}
    \def \x{8}
    \def \s{0.02}

     \draw [fill=gray!15,draw=none] (-0.05,0.75+2.5*\xs) rectangle (5.4,0+0.25*\xs);   
     \draw [fill=blue!15,draw=none] (-0.05,-0.25*\xs) rectangle (5.4,-0.75-2.5*\xs);       
    
    \draw (-0.1,0.45) -- (5,0.45);   
    \draw (-0.1,0.75) -- (5,0.75);  
    \draw (-0.1,0.15) -- (5,0.15);
    \draw (-0.1,-0.15) -- (5,-0.15);
    \draw (-0.1,-0.45) -- (5,-0.45);   
    \draw (-0.1,-0.75) -- (5,-0.75);   
    \draw [fill=GreenYellow,draw=black] (1,0.15+\xs) rectangle (1+\b,-0.15-\xs); 
    \draw [fill=GreenYellow,draw=black] (3,0.15+\xs) rectangle (3+\b,-0.45-\xs); 
    \draw [fill=GreenYellow,draw=black] (4,0.75+\xs) rectangle (4+\b,-0.15-\xs); 
    \node at (1+\b/2+0.01,0) {\footnotesize{$U_{\!1}$}};
    \node at (3+\b/2+0.01,-0.15) {\footnotesize{$U_{\!2}$}};
    \node at (4+\b/2+0.01,0.3) {\footnotesize{$U_{\!3}$}};    
    
    \draw [fill=white,draw=black] (0.5,0.75+\xs) rectangle (0.5+\b,0.45-\xs);
    \draw [fill=white,draw=black] (0.0,0.15+\xs) rectangle (0.0+\b,0.15-\xs);   
    \draw [fill=white,draw=black] (0.0,-0.15+\xs) rectangle (0.0+\b,-0.45-\xs);
    \draw [fill=white,draw=black] (0.5,-0.45+\xs) rectangle (0.5+\b,-0.75-\xs);
    
    \draw [fill=white,draw=black] (1,0.75+\xs) rectangle (1+\b,0.75-\xs);
    \draw [fill=white,draw=black] (1.5,0.45+\xs) rectangle (1.5+\b,0.15-\xs);  
    \draw [fill=white,draw=black] (1.5,-0.15+\xs) rectangle (1.5+\b,-0.45-\xs);
    \draw [fill=white,draw=black] (2,-0.75+\xs) rectangle (2+\b,-0.75-\xs);
    \draw [fill=white,draw=black] (2,0.75+\xs) rectangle (2+\b,0.15-\xs);

    \draw [fill=white,draw=black] (2.5,-0.45+\xs) rectangle (2.5+\b,-0.75-\xs);  
    
    \draw [fill=white,draw=black] (3,0.75+\xs) rectangle (3+\b,0.45-\xs); 
    
    \draw [fill=white,draw=black] (3.5,0.45+\xs) rectangle (3.5+\b,0.15-\xs);
    \draw [fill=white,draw=black] (3.5,-0.45+\xs) rectangle (3.5+\b,-0.75-\xs);

    \draw [fill=white,draw=black] (4.5,-0.15+\xs) rectangle (4.5+\b,-0.45-\xs);
    \draw [fill=white,draw=black] (4.5,0.75+\xs) rectangle (4.5+\b,0.45-\xs);  
    
    \draw [fill=white,draw=black] (5,0.15+1.6*\xs) rectangle (5+\b,0.15-1.6*\xs);
    \draw [fill=white,draw=black] (5,0.45+1.6*\xs) rectangle (5+\b,0.45-1.6*\xs); 
    \draw [fill=white,draw=black] (5,0.75+1.6*\xs) rectangle (5+\b,0.75-1.6*\xs); 
    \draw [fill=white,draw=black] (5,-0.15+1.6*\xs) rectangle (5+\b,-0.15-1.6*\xs);
    \draw [fill=white,draw=black] (5,-0.45+1.6*\xs) rectangle (5+\b,-0.45-1.6*\xs); 
    \draw [fill=white,draw=black] (5,-0.75+1.6*\xs) rectangle (5+\b,-0.75-1.6*\xs);               
     \draw[thin] (5+\b-0.05,0.15-0.07) arc (0:180:0.1);
     \draw[thin,->] (5+\b-0.05-0.11,0.15-0.07) -- (5+\b-0.05,0.15-0.07+0.15); 
     \draw[thin] (5+\b-0.05,0.45-0.07) arc (0:180:0.1);
     \draw[thin,->] (5+\b-0.05-0.11,0.45-0.07) -- (5+\b-0.05,0.45-0.07+0.15);   
     \draw[thin] (5+\b-0.05,0.75-0.07) arc (0:180:0.1);
     \draw[thin,->] (5+\b-0.05-0.11,0.75-0.07) -- (5+\b-0.05,0.75-0.07+0.15);     
     \draw[thin] (5+\b,0.15+\s) -- (5+\b+7*\s,0.15+\s);
     \draw[thin] (5+\b,0.15-\s) -- (5+\b+7*\s,0.15-\s);     
     \draw[thin] (5+\b,0.45+\s) -- (5+\b+7*\s,0.45+\s);
     \draw[thin] (5+\b,0.45-\s) -- (5+\b+7*\s,0.45-\s);   
     \draw[thin] (5+\b,0.75+\s) -- (5+\b+7*\s,0.75+\s);
     \draw[thin] (5+\b,0.75-\s) -- (5+\b+7*\s,0.75-\s);   

     \draw[thin] (5+\b-0.05,-0.15-0.07) arc (0:180:0.1);
     \draw[thin,->] (5+\b-0.05-0.11,-0.15-0.07) -- (5+\b-0.05,-0.15-0.07+0.15); 
     \draw[thin] (5+\b-0.05,-0.45-0.07) arc (0:180:0.1);
     \draw[thin,->] (5+\b-0.05-0.11,-0.45-0.07) -- (5+\b-0.05,-0.45-0.07+0.15);   
     \draw[thin] (5+\b-0.05,-0.75-0.07) arc (0:180:0.1);
     \draw[thin,->] (5+\b-0.05-0.11,-0.75-0.07) -- (5+\b-0.05,-0.75-0.07+0.15);     
     \draw[thin] (5+\b,-0.15+\s) -- (5+\b+7*\s,-0.15+\s);
     \draw[thin] (5+\b,-0.15-\s) -- (5+\b+7*\s,-0.15-\s);     
     \draw[thin] (5+\b,-0.45+\s) -- (5+\b+7*\s,-0.45+\s);
     \draw[thin] (5+\b,-0.45-\s) -- (5+\b+7*\s,-0.45-\s);   
     \draw[thin] (5+\b,-0.75+\s) -- (5+\b+7*\s,-0.75+\s);
     \draw[thin] (5+\b,-0.75-\s) -- (5+\b+7*\s,-0.75-\s);       
                          
%%%%%%%%%%%%%%%%%%%%%%%%%%%%%%%%%

%\node at (5/2+\b/2-0.2/2+\x/2,0) {$\implies$};

%%%%%%%%%%%%%%%%%%%%%%%%%%%%%%%%
     \draw [fill=gray!15,draw=none] (-0.05+\x,0.75+2.5*\xs) rectangle (5.4+\x,0+0.25*\xs);   
     \draw [fill=blue!15,draw=none] (-0.05+\x,-0.25*\xs) rectangle (5.4+\x,-0.75-2.5*\xs);       
    
    \draw (-0.1+\x,0.45) -- (5+\x,0.45);   
    \draw (-0.1+\x,0.75) -- (5+\x,0.75);  
    \draw (-0.1+\x,0.15) -- (5+\x,0.15);
    \draw (-0.1+\x,-0.15) -- (5+\x,-0.15);
    \draw (-0.1+\x,-0.45) -- (5+\x,-0.45);   
    \draw (-0.1+\x,-0.75) -- (5+\x,-0.75);   
    \draw [fill=GreenYellow,draw=black] (1+\x,0.15+\xs) rectangle (1+\x+\b,0.15-\xs); 
    \draw [fill=GreenYellow,draw=black] (1+\x,-0.15+\xs) rectangle (1+\x+\b,-0.15-\xs);     
    \draw [fill=GreenYellow,draw=black] (3+\x,0.15+\xs) rectangle (3+\x+\b,0.15-\xs); 
    \draw [fill=GreenYellow,draw=black] (3+\x,-0.15+\xs) rectangle (3+\x+\b,-0.45-\xs);     
    \draw [fill=GreenYellow,draw=black] (4+\x,0.75+\xs) rectangle (4+\x+\b,0.15-\xs); 
    \draw [fill=GreenYellow,draw=black] (4+\x,-0.15+\xs) rectangle (4+\x+\b,-0.15-\xs);     
    
    \draw [fill=white,draw=black] (0.5+\x,0.75+\xs) rectangle (0.5+\b+\x,0.45-\xs);
    \draw [fill=white,draw=black] (0.0+\x,0.15+\xs) rectangle (0.0+\b+\x,0.15-\xs);   
    \draw [fill=white,draw=black] (0.0+\x,-0.15+\xs) rectangle (0.0+\b+\x,-0.45-\xs);
    \draw [fill=white,draw=black] (0.5+\x,-0.45+\xs) rectangle (0.5+\b+\x,-0.75-\xs);
    
    \draw [fill=white,draw=black] (1+\x,0.75+\xs) rectangle (1+\b+\x,0.75-\xs);
    \draw [fill=white,draw=black] (1.5+\x,0.45+\xs) rectangle (1.5+\b+\x,0.15-\xs);  
    \draw [fill=white,draw=black] (1.5+\x,-0.15+\xs) rectangle (1.5+\b+\x,-0.45-\xs);
    \draw [fill=white,draw=black] (2+\x,-0.75+\xs) rectangle (2+\b+\x,-0.75-\xs);
    \draw [fill=white,draw=black] (2+\x,0.75+\xs) rectangle (2+\b+\x,0.15-\xs);

    \draw [fill=white,draw=black] (2.5+\x,-0.45+\xs) rectangle (2.5+\b+\x,-0.75-\xs);  
    
    \draw [fill=white,draw=black] (3+\x,0.75+\xs) rectangle (3+\b+\x,0.45-\xs); 
    
    \draw [fill=white,draw=black] (3.5+\x,0.45+\xs) rectangle (3.5+\b+\x,0.15-\xs);
    \draw [fill=white,draw=black] (3.5+\x,-0.45+\xs) rectangle (3.5+\b+\x,-0.75-\xs);

    \draw [fill=white,draw=black] (4.5+\x,-0.15+\xs) rectangle (4.5+\b+\x,-0.45-\xs);
    \draw [fill=white,draw=black] (4.5+\x,0.75+\xs) rectangle (4.5+\b+\x,0.45-\xs);  
    
    \draw [fill=white,draw=black] (5+\x,0.15+1.6*\xs) rectangle (\x+5+\b,0.15-1.6*\xs);
    \draw [fill=white,draw=black] (\x+5,0.45+1.6*\xs) rectangle (\x+5+\b,0.45-1.6*\xs); 
    \draw [fill=white,draw=black] (\x+5,0.75+1.6*\xs) rectangle (\x+5+\b,0.75-1.6*\xs); 
    \draw [fill=white,draw=black] (\x+5,-0.15+1.6*\xs) rectangle (\x+5+\b,-0.15-1.6*\xs);
    \draw [fill=white,draw=black] (\x+5,-0.45+1.6*\xs) rectangle (\x+5+\b,-0.45-1.6*\xs); 
    \draw [fill=white,draw=black] (\x+5,-0.75+1.6*\xs) rectangle (\x+5+\b,-0.75-1.6*\xs);               
     \draw[thin] (\x+5+\b-0.05,0.15-0.07) arc (0:180:0.1);
     \draw[thin,->] (\x+5+\b-0.05-0.11,0.15-0.07) -- (\x+5+\b-0.05,0.15-0.07+0.15); 
     \draw[thin] (\x+5+\b-0.05,0.45-0.07) arc (0:180:0.1);
     \draw[thin,->] (\x+5+\b-0.05-0.11,0.45-0.07) -- (\x+5+\b-0.05,0.45-0.07+0.15);   
     \draw[thin] (\x+5+\b-0.05,0.75-0.07) arc (0:180:0.1);
     \draw[thin,->] (\x+5+\b-0.05-0.11,0.75-0.07) -- (\x+5+\b-0.05,0.75-0.07+0.15);     
     \draw[thin] (\x+5+\b,0.15+\s) -- (\x+5+\b+7*\s,0.15+\s);
     \draw[thin] (\x+5+\b,0.15-\s) -- (\x+5+\b+7*\s,0.15-\s);     
     \draw[thin] (\x+5+\b,0.45+\s) -- (\x+5+\b+7*\s,0.45+\s);
     \draw[thin] (\x+5+\b,0.45-\s) -- (\x+5+\b+7*\s,0.45-\s);   
     \draw[thin] (\x+5+\b,0.75+\s) -- (\x+5+\b+7*\s,0.75+\s);
     \draw[thin] (\x+5+\b,0.75-\s) -- (\x+5+\b+7*\s,0.75-\s);   

     \draw[thin] (\x+5+\b-0.05,-0.15-0.07) arc (0:180:0.1);
     \draw[thin,->] (\x+5+\b-0.05-0.11,-0.15-0.07) -- (\x+5+\b-0.05,-0.15-0.07+0.15); 
     \draw[thin] (\x+5+\b-0.05,-0.45-0.07) arc (0:180:0.1);
     \draw[thin,->] (\x+5+\b-0.05-0.11,-0.45-0.07) -- (\x+5+\b-0.05,-0.45-0.07+0.15);   
     \draw[thin] (\x+5+\b-0.05,-0.75-0.07) arc (0:180:0.1);
     \draw[thin,->] (\x+5+\b-0.05-0.11,-0.75-0.07) -- (\x+5+\b-0.05,-0.75-0.07+0.15);     
     \draw[thin] (\x+5+\b,-0.15+\s) -- (\x+5+\b+7*\s,-0.15+\s);
     \draw[thin] (\x+5+\b,-0.15-\s) -- (\x+5+\b+7*\s,-0.15-\s);     
     \draw[thin] (\x+5+\b,-0.45+\s) -- (\x+5+\b+7*\s,-0.45+\s);
     \draw[thin] (\x+5+\b,-0.45-\s) -- (\x+5+\b+7*\s,-0.45-\s);   
     \draw[thin] (\x+5+\b,-0.75+\s) -- (\x+5+\b+7*\s,-0.75+\s);
     \draw[thin] (\x+5+\b,-0.75-\s) -- (\x+5+\b+7*\s,-0.75-\s);

     \draw [fill=gray!15,draw=none] (0.1,1+0.2) rectangle (0.1+0.25,1+0.25+0.2);
     \draw [fill=blue!15,draw=none] (1.6,1+0.2) rectangle (1.6+0.25,1+0.25+0.2);     
     \node[gray] at (0.55,1.325) {\footnotesize{$\bar A$}};
     \node[blue] at (2.05,1.325) {\footnotesize{$\bar B$}};
   
    \end{tikzpicture}
    \caption{\textbf{Circuit Cutting Example.} The circuit on the left has 3 gates which cross a partition of the qubits into 2 clusters, $\bar{A}$ and $\bar{B}$, indicated by the background colors. Using a quasi-probability decomposition of each gate, it becomes possible to express the output of the circuit on the left as a sum of outputs of circuits having the form on the right, where each of the indicated gates is replaced gates which act locally on qubits in each cluster. The total number of circuits needed (the \textit{sampling overhead} depends on specific properties of each gate being cut, along with the particular details of the cutting method used. NOTE: Image is taken from \cite{Piveteau_2023}, and is used with permission.}
    \label{fig:cutting}
\end{figure}

Of these 3 categories, circuit knitting for decomposition (aka, ``circuit cutting") has been the most actively explored category in the literature. We focus on that category in the remainder of this sub-section, and show in Figure \ref{fig:cutting} an example of circuit cutting. The basic idea of circuit cutting was first introduced in \cite{Bravyi_2016-knitting}, which showed that sparse circuits of width $n+k$ could be simulated using quantum computers with $n$ qubits and classical post-processing, where the simulation time required scaled as $2^{\mathcal{O}(k)}*\mathrm{poly}(n)$. Since then, two additional lines of work have emerged. One method, introduced in \cite{PhysRevLett.125.150504}, uses a decomposition of the relevant tensors across the partition to generate a set of state preparations and measurements which are inserted into the circuit at the location of the cut. The overhead of this method is exponential in the number of cuts. Another,  introduced in \cite{Piveteau_2023}, the gate being cut is randomly replaced by one of those from a \textit{quasi-probabilility decomposition} of the gate. (A quasi-probability decomposition is one where the coefficients of the decomposition can be negative.) The overhead is also exponential in the number of cuts. Interestingly, for some cutting problems based on this method, the use of classical communication does help \cite{Piveteau_2023,brenner2023optimal}, whereas for others, it does not \cite{schmitt2023cutting}.

Some implementations of circuit cutting have shown improvements in the accuracy of the results obtained with the method as compared to running the original circuit on a larger-qubit-count computer \cite{Ayral_2020,Tang_2021-cutqc,Ying_2023, Bechtold_2023}. Combined with improvements in the underlying hardware, these implementations validate the usefulness of circuit cutting for realizing high-accuracy implementations of large-width circuits. In addition, implementations of cutting have shown how to reduce the risk of barren plateaus in variational algorithms \cite{Tuysuz2023}, improve circuit compilation and runtime performance \cite{brandhofer2023optimal}, and enable new approaches for simulating highly-correlated spin chains \cite{gentinetta2023overheadconstrained}. In addition, several software packages have been developed to support the use of circuit cutting. These include CutQC \cite{tang2022cutting}, ScaleQC \cite{tang2022scaleqc}, and the Qiskit Circuit Knitting Toolbox \cite{circuit-knitting-toolbox}.

Addressing the exponential overhead associated with circuit cutting is an active area of research. Ideas being explored include using randomly-inserted state preparations and measurements \cite{Lowe_2023}, approximate cutting \cite{chen2022approximate, chen2023efficient}, and ancilla-free methods \cite{pednault2023alternative,harada2023doubly}.

Overall, circuit knitting protocols can enable current and near-future quantum computers to run circuits at a scale (and depth) larger than what they would natively allow given the number of physical qubits available. As such, circuit knitting is a powerful approach for emulating larger-scale (and higher-quality) quantum computers with currently-available systems.

These recent developments of variational algorithms, error mitigation, and circuit knitting lay a foundation for running larger-sized circuits than what hardware could natively support, and help find approximate, ``good enough" circuits for problems. Having discussed these metods, we turn next to the question of the kinds of ``commercial problems" for which quantum computing could make an impact, and what the private sector has done with access to quantum computers and quantum computing expertise.

\subsubsection{Commercial Exploration and Adoption of Quantum Computers}
\label{sec:commercialqc}
As noted in Section \ref{sec:backdrop}, the private sector has taken a keen interest in quantum computing. Many companies have created teams to explore the potential uses of quantum computers to business-relevant problems. In this sub-section, we review and summarize some of the predominant themes from the body of literature which has resulted from commercial end-user activity around quantum computing\footnote{Note  there is a wide variety of literature on the topics below which touches on similar or adjacent matters, but was not necessarily pursued in -- or directed toward -- a commercial/practical context. We omit such literature and references here, lest our bibliography expand to one of biblical proportions. We invite the reader to consider \cite{Bharti_2022},  \cite{dalzell2023quantum}, and \cite{auyeung2023quantum} as useful resources for such information.}. We emphasize that our purpose here is not to make any claims regarding the attainment of quantum advantage in any of these industries, nor to argue that the topics considered by the teams within those industries are the definitive ones which will provide commercial value. Instead, we seek to simply inform the reader of the predominant themes which have emerged as the private sector has begun engaging with quantum computing, and refer the reader to Ref. \cite{hoefler2023disentangling} for a perspective on where quantum advantage is most likely to be obtained. 

We consider the intersection of broad ``quantum application areas" and specific industry segments/verticals. The application areas can be broken down into 3 categories: simulating nature, data processing (including machine learning), and operations research (which includes financial engineering, search, and optimization). There are a wide variety of industries investigating quantum computing; thus, for each combination of application area and industry vertical, there are many possibilities as to how quantum computing could make an impact. In addition, a quantum algorithm developed for one particular industry or application very often can be used across others. Although some amount of application or use-case tailoring is necessary, a given quantum algorithm finds wide applicability.

We focus our review on 3 example industries which have been most active in exploring the promise of quantum computing: finance, aerospace/automotive, and high-technology manufacturing. Collectively, the value at stake within these industries may exceed on the order of hundreds of billions of dollars over the next decade \cite[Exhibit 4]{mk-apps}.

Although the idea of quantum computing was first developed around the application area of simulating nature, an important industry where quantum computing is explored is finance. The abstract of a recent survey paper by JP Morgan Chase \cite{herman2022survey} states ``Quantum computers are expected to ... have transformative impact on numerous industry sectors, particularly finance."

This sentiment was further echoed by work from Hyperion Research \cite{sorenson22} which, based off of survey results of companies in the quantum space, suggest the finance sector will be one of the largest users of quantum computation by 2025. Numerous financial firms are exploring the use of quantum computing \cite[Table 1]{ukfinance}. Areas being investigated by them include risk management \cite{Rebentrost_2018,Woerner_2019,Stamatopoulos_2020,zhu2022copulabased,leclerc2022financial,stamatopoulos2022towards, cherrat2023quantum,gomez2022survey,chakrabarti2021threshold,egger2020quantum,herman2022survey,stamatopoulos2023derivative,ghosh2023energy,o2023quadratic,Herman_2023,stamatopoulos2023derivative,ghosh2023energy,Alcazar_2022,tang2022quantum,wang2023option}, customer segmentation and other machine learning problems \cite{pistoia2021quantum,orus2019quantum,Herman_2023,schetakis2023quantum}, portfolio optimization \cite{egger2020quantum,herman2022survey,vesely2023finding,mattesi2023financial,han2022quantum,chen2023quantum,dalzell2023quantum}, and fraud detection \cite{9915517,suzuki2023quantum,kyriienko2022unsupervised,Zoufal_2023,innan2023financial,innan2023financial-2}.  For further reference, Tables 5 and 6 of \cite{egger2020quantum}, Table 1 of \cite{Herman_2023}, and Table 1 of \cite{albareti2022structured} show how various quantum algorithms can be used for risk management, optimization, and machine learning in finance, and provides the relevant financial services which are impacted by these approaches.

The aerospace/automotive industry is another exploring the adoption of quantum computing \cite{mk-auto,bayerstadler2021industry}. Several major companies in these industries have launched ``quantum challenges" \cite{AirbusChallenge,Airbus-BMW-Challenge,AirbusChallengeFinalists,BMWChallenge,BMWChallengeWinners} to generate research and activity around potential use cases. In doing so, these companies are stimulating research in areas of key concern to their business, and driving awareness in the quantum computing industry and community of these problems:

\begin{quote}
    “We organised this competition to make a bridge to the quantum computing community,” ... “And the response from the community was enthusiastic! We believe this challenge can serve as a new template for how businesses like Airbus can link with quantum computing researchers across the globe to transform this fundamental research topic into an impactful computing solution for a wide range of industrial applications.” \cite{AirbusChallengeFinalists}
\end{quote}

Further, companies have conducted their own assessments of business-relevant problems for which additional work is required \cite[Table 1]{luckow2021quantum}. Specific topics in aerospace which have been investigated include materials corrosion \cite{gujarati2023quantum}, aircraft design \cite{fuller2021approximate}, and airline logistics/planning \cite{mohammadbagherpoor2021exploring,pilon2021aircraft,consul2023quantum,makhanov2023quantum,consulpacareu2023quantum,chai2023finding}, and computational fluid dynamics \cite{lapworth2022hybrid,lapworth2022implicit}. In the automotive industry, topics explored include predictive process monitoring \cite{hill2023intercase}, supply chain logistics optimization \cite{correll2022quantum,Sanches_2022}, batteries \cite{Rice_2021,Motta_2020,Kim_2022,gujarati2021heuristic,Takeshita_2020,Shokrian_Zini_2023,Delgado_2022}, catalysis/materials \cite{shirai2023computational,amsler2023quantumenhanced,Omiya_2022}, and several others \cite{bentley2022quantum,Streif_2021,dollen2023predicting,yarkoni2021solving,cattelan2022modeling,chatterjee2023hybrid,mckinsey-mobility}.

Finally, the high-technology manufacturing and materials and chemistry industry will benefit from quantum computing. Here, the primary application of quantum computation is simulating the behavior of novel molecules and materials \cite{reiher2017elucidating,elfving2020will,greene2022modelling,di2023applicability,dipaola2023applicability,Liu2022,bauer2020quantum,von2021quantum}, though data processing (in the form of generative modeling \cite{kao2023exploring,zoufal2021generative}) and operations improvements \cite{amaro2022case} are also promising. For this industry, it is anticipated it may benefit in the near-term from the fact that problems in chemistry/materials generally require a comparative small amount of classical data as part of the quantum computation (typically, a description of the molecule or system to be simulated). Such ``big (quantum) compute, small data" problems have been identified as the most promising ones to consider for realizing useful quantum computing in the near-term \cite{hoefler2023disentangling,Rossmannek_2021,rossmannek2023quantum}. What's more, this industry has been recognized as one of the top industries for finding end-use applicability of quantum computers \cite{hyperion-2023}.

Beyond those industries described above, substantial work has taken place exploring the use of quantum computing in many other industrial sectors, including biosciences, lifesciences \& healthcare \cite{outeiral2021prospects,emani2021quantum,blunt2022perspective,flother2023state,basu2023towards,santagati2023drug,Emani_2021,Blunt_2022,Cordier_2022,zinner2021quantum,london2023peptide,doga2023perspective}, telecommunications \cite[Table 2]{martin2021quantum} \cite{amoretti2022classical,pabst2022quantum,phillipson2023quantum}, logistics \cite{mk-apps,ibm-ibv-logistics,osaba2022systematic,harwood2021formulating,gachnang2022quantum}, and the energy/utilities sector \cite{golestan2023quantum,pabst2022quantum,9831167,ajagekar2019quantum,berger2021quantum,sævarsson2023stochastic,jong2023quantum,o2023quadratic,ghosh2023energy,colucci2022power,vandelli2023theory,sagingalieva2023photovoltaic}.

In total, quantum computing appears to be likely to impact many computational problems across a variety of industrial sectors. Additional exploration is needed to further identify and refine the most promising use cases, assess the practicality of quantum advantage for those use cases, and understand how to integrate quantum computing into industry-relevant workflows. Much has been written on how the private sector can prepare for and embrace quantum computing; see \cite{ibm-ibv-q-ready,bcq-business-ready, mk-apps, hbr-quantum,mit-quantum,deloitte-quantum,Saltan2022} and \cite{andreas2021industry,bova2021commercial,gupta2023effects} for further ideas and references. We emphasize that commercial entities are vital partners for realizing the full commercial impact of quantum computing, and private sector investments and engagement is crucial. A recent analysis of the levels of investment and engagement are encouraging in this regard \cite{bcq-business-ready}.

In sum, further research in the application of quantum computers for commercial problems continues, and the possibility exists that some as-yet-unexplored set of problems exist which are of importance, but which also could be tackled with the kinds of quantum computers that will become available over the next several years. In addition, further advances in hardware error rates, variational algorithms, and error mitigation should also move closer to the present the time horizon at which quantum computers provide commercial value.

Although the commercial value at stake is large, and the private sector has made substantial strides in exploring the use of quantum computers, realizing near-term commercial value has remained challenging. The remainder of this section provides a discussion on what a path ahead could look like for quantum computing to make an impact for national prosperity.

\subsection{Realizing Near-Term Value of Quantum Computers}
\label{sec:near-term-value-science}

The past decade has given rise to a tremendous increase in the exploration of quantum algorithms for a full range of hard problems. Resource estimates for quantum algorithms have dropped substantially, though the resources implied by many current estimates remain out of reach for the foreseeable future. As a result, it is becoming clear that variational (or post-variational) algorithms offer the most promise for realizing applications of quantum computers in the near term. Combining these algorithms with the use of error suppression and/or mitigation techniques (and possibly some very limited forms of error correction) will likely create scientific and economic value from new quantum hardware. And with more robust quantum computers, these algorithms will make further progress.

In addition, as noted in the discussion of Figure \ref{fig:circuit-resource-estimate}, there is a gap in the research literature as to problems which could be addressed by circuits approximately $1K$ qubits in width and requiring  $\sim 1-10K$ $T$ gates. Further research would help clarify what useful applications (if any) exist in that region, and also potentially help rule out the possibility of a cryptographically-relevant threat residing in that space as well. In addition, the use of circuit knitting techniques to break apart large-sized circuits into smaller-sized ones may extend the reach of the applications addressable by current and near-future quantum computers.

One of the challenges in assessing time frames over which quantum computing attains an advantage over classical is that advances in classical methods can ``narrow the gap", so speak, between quantum and classical compute. Examples of this over the last few years have been discussed in the previous section. The core argument for beyond-classical experiments has been showing that a quantum computer could perform some task using fewer resources (usually, time) for which the estimated classical resources are astronomically prohibitive to attain or use in a reasonable amount of time. However, new classical methods continue to be developed \cite{bravyi2023classical,PhysRevLett.128.220503,Bravyi_2016,Pashayan_2022,Bravyi_2016,Qassim_2021,gosset2021fast}, especially for simulating state-of-the-art quantum computing experiments  \cite{villalonga2020establishing,pan2021simulating,huang2020classical,pednault2019leveraging,tindall2023efficient, begušić2023fast, kechedzhi2023effective, anand2023classical,huang2021efficient,Wu_2018,pan2023efficient}, and ``quantum-inspired" algorithms have been proposed \cite{gilyen2018quantuminspired,Tang_2021,Tang_2019,Gilyen_2022,Gharibian_2022,shao2023faster,bakshi2023improved,gall2023robust,gourianov2022quantum}. For a survey of classical simulation methods, see \cite{xu2023herculean,Bridgeman_2017,orus2014practical,young2023simulating}. As a result, a kind of ``friendly competitive escalation" has developed between the quantum and classical computing communities (however, the methods being explored on the classical simulation side are \textit{approximate}, not exact or brute-force in nature).

While the back-and-forth between classical versus quantum computation will probably continue for a while, how the research community has grappled with the question of defining and attaining quantum advantage offers a powerful lesson as quantum begins to impact the commercial sector; namely, that it should be expected that the first demonstrations of the application of quantum computing to practical problems (either of scientific or commercial value) will be accompanied by both investigations of classical methods, and very likely follow-on work of both quantum and classical research teams. In this sense, the attainment of quantum advantage -- however that ends up being defined-- should be considered as being defined \textit{less} by one particular experiment or result, and \textit{more} by a general transition taking place over an extended time period. It is reasonable to expect that back-and-forth to continue for a while, and focusing on the general trends and relationship between quantum and classical computing will be more important than, e.g., fixating on any one particular advance or announcement.

However, it is reasonable to expect that the near-term value of quantum computers can also be realized outside of the commercial space, and for problems where the scientific community has developed clearer guideposts to help indicate where near-term quantum computers could make an impact. Harkening back to quotations from Richard Feynman in the opening of this paper, these problems will most likely relate to the topic of simulating the behavior of Nature.

One particular problem singled out is simulating the time evolution (dynamics) of a particular family of spin systems; namely, a one-dimensional (1D) Heisenberg spin system with a random magnetic field \cite{childs2018toward}. (For other families of spin systems, simulating their dynamics has also been proposed as a possible path to demonstrating beyond-classical quantum computation \cite{Flannigan_2022,Babbush_2018,Nam_2019}.)  Spin systems can exhibit strong correlations amongst the spins, which may make classical simulation of large spin systems (e.g., hundreds of spins) intractable \cite{pearson2020simulating,Wahl17,luitz2015many,Flannigan_2022}. This is because the primary method by which those classical simulations are done involves brute-force simulation of the dynamics, which incurs an unfavorable exponential resource scaling. (Though approximate methods may be applied.) Further, until recently, using quantum computers to simulate spin systems was also restricted to relatively small systems; namely, on the order of tens of spins \cite{smith2019simulating,sun2021quantum,yeter2021scattering,francis2020}.

However, as the previous section noted, recent work has extended the frontier of using quantum computing to simulate two-dimensional (2D) spin models \cite{q-utility-23}. This work shows that, for a quantum circuit inspired by a particular class of spin models, it is possible to use error mitigation to estimate properties in regimes where brute-force classical approximation methods break down. (That said, it is possible to use other, non-brute-force methods to attain comparable results, both amongst the classical methods and the quantum error-mitigated ones \cite{anand2023classical, beguvsic2023fast,liao2023simulation,tindall2023efficient,kechedzhi2023effective}. As noted above, it is reasonable to expect back-and-forth between advances in classical simulation methods and quantum error mitigation advances. Hence, the results in \cite{q-utility-23} cannot be taken as evidence of quantum advantage, per se, but they highlight how using circuits which are tailored to the connectivity of the hardware, along with quantum error mitigation techniques and improvements in the hardware itself, could enable quantum computers to be useful in the near-term. In addition, they exemplify how in the near-term, it may be possible to realize non-trivial scientific applications of quantum computers. As such, there are grounds for a cautious optimism that near-term quantum computers will be useful for scientific exploration and research. Of course, whether and how the ideas developed for those purposes can be re-used or adjusted for commercial purposes remains to be seen, and is an active area of engagement between quantum computing providers and end-users \cite{ibm-100x100,ibm-hep-twg,basu2023towards}.

In sum, circuit knitting, variational algorithms, and error mitigation offer new approaches for realizing near-term value from quantum computers. It is possible to use these approaches for both current and future quantum computers. The commercial sector has explored how to use quantum computers for business benefit, especially the finance, aerospace/automotive, and high-technology manufacturing industries. In parallel, the scientific community has identified one particular class of scientifically-relevant problems (simulating the behavior of spin systems) for which quantum computers would enable marked advances in our understanding of the behavior of Nature. The ideas and techniques being developed for tackling such problems at classically-intractable scales should have relevance for unlocking value in a commercial context, though translating them will require collaboration between private sector end-users and commercial providers of access to quantum computers.

This section primarily discussed the implications of quantum computing for national prosperity. However, ever since the development of Shor's algorithm in 1994, it has been known that quantum computing poses a threat to cybersecurity (and by extension, national security). Having a solid handle on this threat is essential for informed discussion on the benefits and risks of quantum computers. We turn to the topic of cybersecurity and quantum computing in the next section.

\section{Implications of Quantum Computing for National Security}
\label{sec:national security}
\addtocontents{toc}{%
  \smallskip\protect\parbox[t]{\textwidth}{\textit{Surveys what is known about the uses of quantum computers for cryptanalysis, and analyzes the circuits sizes needed for cryptographically-relevant quantum computing.}}\par~\\}

Having discussed the implications of quantum computers for commerce and science, we turn in this section to discussing the implications for cybersecurity. We begin with a brief overview of public-key cryptosystems, discuss common quantum algorithms for breaking those cryptosystems, and then present conservative estimates (based on existing literature) on the typical error rates a quantum computer would need to break cryptosystems with varying numbers of bits of security. The reader is also invited to look at \cite[Section 4]{NAP25196} and \cite{mavroeidis2018impact} for additional discussion on this topic.

\subsection{An Overview of Public-Key Cryptosystems}
\label{sec:pki}
Public-key cryptography involves techniques for encrypting (or signing) data using pairs of related keys, consisting of a public key and a private key.  As its name indicates, the public key is assumed to be public information.  In a public-key encryption system, anybody with the public key can encrypt a message to produce the ciphertext, while the recipient uses their private key to decrypt it, recovering the original message.  Similarly, in a public-key signature system, the author of a document uses their private key to digitally sign a message, which anybody with the public key can verify, providing authentication.  Public-key cryptography is also sometimes known as \textit{asymmetric} cryptography -- the public and private keys are different -- in contrast with \textit{symmetric-key}, involving a single secret key, which is used to both encrypt and decrypt.

The primary threat quantum computers pose to existing cryptosystems is attacking asymmetric cryptography. For symmetric-key cryptosystems, the relevant quantum attack is Grover’s search algorithm \cite{grover1996fast}, which would speed up (by a quadratic factor) a brute-force search for the key. However, this attack can be resolved by doubling the key length for the impacted algorithms. In fact, there are mitigating factors suggesting  Grover’s algorithm will not speed up brute force key search as dramatically as one might suspect: it was proven in 1997 \cite{PhysRevA.60.2746} that in order to obtain the full quadratic speedup, all the steps of Grover’s algorithm must be performed in series. In the real world, where attacks on cryptography use massively parallel processing, the practical advantage of using Grover’s algorithm for symmetric-key cryptography will be substantial smaller \cite{cryptoeprint:2017/811}. While other attacks on symmetric-key cryptosystems exist \cite{chailloux2017efficient,kaplan2016breaking,santoli2017using}, the full ramifications of these attacks are currently unknown.

In contrast, the ramifications of quantum computing for public-key cryptosystems are much better understood, and hence, we focus on such cryptosystems in this paper. Various public-key algorithms are widely used today, including RSA \cite{wiki-RSA-cryptosystem}, the Diffie-Hellman key agreement \cite{wiki-Diffie-Hellman}, the digital signature algorithm (DSA) \cite{wiki-DSA}, and elliptic curve cryptography (ECC) algorithms \cite{wiki-ECC}. The mathematics underlying these cryptosystems is different, which has implications for how quantum computers can break them:
\begin{itemize}
    \item RSA is based on arithmetic with integers, and its security is related to the hardness of factoring large integers into their prime factors.
    \item Both Diffie-Hellman and DSA are examples of finite field cryptography (FFC), which is based on modular arithmetic in a finite field.
    \item ECC is based on the elliptic curve discrete logarithm problems, and includes the elliptic curve Diffie-Hellman (ECDH) key exchange \cite{wiki-ECDH}, as well as the elliptic curve digital signature algorithm (ECDSA) \cite{wiki-ECDSA}.
\end{itemize}

The way a quantum computer would break these cryptosystems is through the use of Shor's algorithm \cite{shor1994algorithms}. This algorithm has several variants: a \emph{discrete logarithm} version, which can be used to break ECC and FCC cryptosystems, and a \emph{factoring} version, which can be used to break RSA. As we'll see below, these variants require different quantum compute resources to break the cryptosystem they attack. If a large-scale fault-tolerant quantum computer is built, FFC, RSA, and ECC algorithms will all be vulnerable to attacks by quantum computers.

Both the National Institute of Standards and Technology (NIST) and the National Security Agency (NSA) have a long history in making and updating recommendations for the various parameters of these cryptosystems to ensure an overall appropriate level of security in light of advances (both classical and quantum) for attacking them. The strength of a cryptographic algorithm is often given in terms of “bits of security.”  For example, common implementations of the Advanced Encryption Standard (AES) \cite{wiki-AES} are AES-128 or AES-256.  The ``-128" refers to AES at the 128-bit security level.  Roughly speaking, this means  about $2^{128}$ operations are required for a brute-force search for the secret key\footnote{For AES, this key is 128 bits long. In public-key cryptosystems, e.g., RSA, an RSA-X key is not necessarily X bits in length.}. For other cryptosystems, the number of bits of security provided by a given key length is different; we provide formulae [Equations \eqref{eq:rsa-bits}, \eqref{eq:ffcdh-bits}, \eqref{eq:ecc-bits}]
in the next Section for them.

Since the first introduction of recommended key sizes by NIST and the NSA, those recommendations have been updated in light of advances in classical compute power and cryptanalysis. We now briefly recall the history of public-key algorithms and their recommended security strengths.  In the early 1990’s, NIST proposed DSA as an approved digital signature algorithm.  DSA had been developed by the NSA, and was being adopted in part due to the high costs of licensing RSA.  Shortly afterwards, an analogous elliptic curve version (ECDSA) was also developed.  Real-world usage of ECC initially remained relatively low, even though ECC has much smaller key sizes.

In the early 2000's, adoption of RSA and ECC increased. In 2002, the broader community recognized that RSA key lengths \cite{wiki-Key-Size} should be at least 1024 bits in length \cite{RSAcost_analysis}\footnote{In comparison, to provide the same level of 80-bit security, ECC keys are only 160 bits.}. In 2005, NIST recommended 80 bits of security would be sufficient through 2010 \cite{SP800-57part1}.  It also suggested 112 bits of security would be sufficient through 2030, corresponding to key lengths of 2048 bits (for RSA) and 224 bits (for ECC). Beyond 2030, the recommendation was for 128 bits of security, corresponding to key lengths of 3072 bits (for RSA) and 256 bits (for ECC).  These general trends have been maintained in the updates NIST has given in recent years \cite{SP800-57part1r1, SP800-57part1r2, SP800-57part1r3}.

In 2012, NIST deprecated the parameter sets for 80 bits of security \cite{SP800-57part1r3}, where ``deprecation" means these parameter sets were disallowed for use except for legacy processing.  NIST also recommended RSA sizes be doubled to 2048 bits and ECC keys be increased to 224 bits, corresponding to 112 bits of security. In subsequent revisions in 2016, NIST provided the same guidance, but with all cryptosystem parameters of 80 bits of security dropped \cite{SP800-57part1r4,SP800-57part1r5}.

In 2005, the NSA announced recommended algorithms for national security systems, collectively known as ``NSA Suite B" \cite{wiki-NSA_Suite_B,NSAsuiteB_factsheet}. NSA Suite B cryptography required the use of 256-bit ECC up to the Secret level and 384 bits for the Top Secret level. Noticeably, it did not authorize the use of RSA nor FFC.  In 2010, the NSA provided updated guidance for Suite B cryptography, and authorized ``during the transition to the use of” ECC that 2048-bit RSA and FFC to be used to protect information up to the Secret level.  No change was made for the Top Secret level. By mid-2014, NSA had updated its guidance to clearly state the use of RSA and FFC would not be allowed after October 1, 2015, even at the Secret level.

In 2015, NSA released a memorandum which clearly stated it was focusing on the effects of quantum computers and the need to shift to post-quantum (also known as quantum-resistant or quantum-safe) cryptography in the near future \cite{CNSSmemo, CNSS_annual_report, CNSSpolicy}. Post-quantum cryptography (PQC) refers to algorithms which are believed to provide protection against attacks from both classical and quantum computers.  The memorandum also stated NSA found it acceptable to continue to use RSA and Diffie-Hellman if the key lengths were a minimum of 3072 bits.  In addition, ECC was allowed to continue to be used with a 384 bit key.  (For an excellent summary of the history and issues raised by this memo, see \cite{koblitz2016}.)  The NSA later published more documents and guidance which led to the demise of NSA Suite B, which was followed by the establishment of the Commercial National Security Algorithm (CNSA) Suite \cite{wiki-CNSA, CNSA, CNSApolicy, CNSAreport}.

We postpone continuing the discussion of post-quantum cryptography to Section \ref{sec:quantum-safe}, and turn now to understanding the threat posed by quantum computers to the public-key cryptosystems which are currently used today. As noted above, all of the public-key cryptosystems used today to protect sensitive data are based on hard problems which are susceptible to being broken using quantum computers. Further, understanding this risk allows for a more thoughtful balance on the matter of weighing the risk against the benefits and opportunities previously discussed in Section \ref{sec:benefits}.

\subsection{Quantum Computing for Breaking Public-Key Cryptography}
\label{sec:q-algs-pki}

In this section, we discuss what is known about using quantum computers to break public-key cryptosystems.  We connect resource estimates for fault-tolerant implementations of algorithms which are known to break existing public-key cryptosystems to both the required capabilities a quantum computer would need to run them as well as the number of bits of security provided by a key of a given cryptosystem and length. Doing so provides a framework for understanding the risks  quantum computers pose to a given cryptosystem using keys of a given level of security.

Our focus on fault-tolerant implementations is motivated by the fact that, to date, no approach based on variational quantum algorithms has been discovered for which a provable (or even plausible) threat exists for current public-key cryptosystems. The variational approaches considered to date consist of 2 parts. First, map the task of attacking a cryptosystem onto a combinatorial optimization problem (using some amount of classical pre-processing to simplify it). Second, solve the resulting optimization problem with variational algorithms. This approach has been used in \cite{anschuetz2019variational,phan2022quantum,jun2023hubo,yan2022factoring, tutul2023shallow,park2023rydbergatom}. Although these approaches have made claims of feasibility (at least in principle) for attacking public-key cryptosystems, implementations on hardware have yielded no conclusive results on the feasibility of scaling them to relevant key sizes \cite{phan2022quantum,jun2023hubo,yan2022factoring,karamlou2021analyzing}. This is because scaling to such key sizes would require circuits which are so large as to be impractical to be run without error correction. In addition, the end-to-end analysis of such approaches lacks the systematic analysis necessary to draw definitive conclusions about their power \cite{grebnev2023pitfalls,khattar2023comment}. Other approaches have been pursued \cite{aizpurua2023hacking,wang2022variational,hegade2023digitizedcounterdiabatic}, but again, there have been no definitive proofs of feasibility for scaling such methods to relevant key sizes. What's more, these algorithms are not proven to be competitive with, e.g., Shor's algorithm, even if they were run in a fault-tolerant manner. Finally, it is currently unknown as to whether these algorithms can offer any speedups in principle relative to using a brute-force search via Grover's algorithm. Hence, we focus here on known fault-tolerant implementations.

Our analysis of these algorithms is based on the observation that while error correction reduces to arbitrary levels the effect of noise, doing so requires increasing the code distance $d_{C}$. Because error correction at finite code distance $d_{C}$ simply \textit{suppresses} the rate of errors, and does not \textit{eliminate} them entirely, there is the possibility an error-corrected quantum computer could fail to successfully run the required circuit for the algorithm. Recall that the relationship between the logical error rate $p_{\mathrm{L}}$, code distance $d_{C}$, code threshold $p_{\mathrm{th}}$ and physical error rate $p_{\mathrm{ph}}$ goes as $p_{\mathrm{L}} \sim (p_{\mathrm{ph}}/p_{\mathrm{th}})^{\mathcal{O}(d_{C})}$. Hence, for fixed code distance $d_{C}$, there is an associated non-zero logical error rate -- \textit{even when the physical error rate is below the code's threshold} -- implying that a fault-tolerant implementation of a given algorithm using a given error-correcting code still has a non-zero probability of failing. 

The required logical error rate necessary to ensure, with high confidence, the algorithm will succeed can be roughly upper-bounded using the framework for analyzing the performance of variational algorithms discussed in Section \ref{sec:benefits}. Namely, if the algorithm requires circuits acting on $q$ qubits and having a depth of $d$, we take $(q*d)^{-1}$ as a loose upper bound on the required value of $p_{\mathrm{L}}$ needed in order to run the algorithm successfully. In turn, this bound places fairly stringent requirements on the underlying physical error rate $p_{\mathrm{ph}}$: it must be sufficiently small so that $p_{\mathrm{L}} < (q*d)^{-1}$. Phrased another way, the physical error rate needs to be small enough such that after error correction on the noisy physical qubits is used, the resulting logical error rate is itself sufficiently small as to enable the error-corrected quantum computer to run the algorithm with a high probability of success.

Our analysis is primarily based on \cite{gidney2021} which currently provides the most efficient quantum algorithm for factoring RSA integers and computing discrete logarithms over a finite field such as the Diffie-Hellman (DH) key agreement and Diffie-Hellman digital signature algorithm (DHDSA). It builds on two papers that provide the foundation for solving the discrete log problem with a quantum computer \cite{haner2020,banegas2021}. (Recall the security of ECC is based on the hardness of solving the discrete log problem.) The former of these two papers \cite{haner2020} effectively uses the techniques in  \cite{gidney2021} to further improve the latter \cite{banegas2021}. These resource estimates are summarized in Table \ref{tab:q-cryptanalysis}. We note for completeness the origin of each of these entries:
\begin{itemize}
    \item Row 1 (RSA - Gidney): See Table 1 of \cite{gidney2021}, with $q$ as ``Abstract Qubits", and $d$ as ``Measurement Depth".
    \item Row 2 (ECC - H\"{a}ner): Formulae given in the final paragraph above Section 10 of \cite{banegas2021}, with $q$ as the number of qubits, and $d$ as the $T$-depth.
    \item Row 3 (ECC - Banegas): See \cite{banegas2021}, with $q$ given as the number of qubits indicated just below Table 4, and $d$ is the number of Toffoli gates, indicated just below Table 6. (From that self-same page, ``...as such the depth is of the same order as the number of TOF [Toffoli] gates." )
\end{itemize}

We emphasize for the reader that the above references are by no means the last and final say regarding resource estimates for cryptanalysis. In particular, we note that just this year, an improved version of Shor's factoring algorithm was released \cite{regev2023efficient}, and already researchers have found ways to improve it \cite{ragavan2023optimizing}. In addition, the algorithm of \cite{regev2023efficient} has been extended to the problem of computing discrete logarithms \cite{ekera2023extending}. We do not include \cite{regev2023efficient} in Table \ref{tab:q-cryptanalysis} and Figure \ref{fig:circuit-resource-estimate} because the resource analysis of this algorithm in terms of Toffoli or $T$ gates has not yet been done. And from \cite{regev2023efficient}: `` It therefore remains to be seen whether the algorithm can lead to improved physical implementations in practice.". Similarly, \cite{litinski2023compute} provides new architectural approaches to implementing Shor's algorithm for ECC, and provides new resource estimates for ECC-256. That said, Table \ref{tab:q-cryptanalysis} does, as far as we authors are aware, provide the state-of-the-art in terms of resource estimates (with analytic formulae) for circuit width and depth for cryptanalysis.

\begin{table}[]
    \centering
    \begin{tabular}{|c|c|c|c|}
    \hline
    \textbf{Cryptographic Scheme} & \textbf{Paper} & \textbf{Circuit Width} ($q$) & \textbf{Required Depth} ($d$)\\ \hline
     RSA    & Gidney \cite{gidney2021} & $3n+.002n\log_{2}n$ & $n^{2}(\log_{2}n + 500)$ \\ \hline
     ECC & H\"{a}ner \cite{haner2020} & $8n+10.2\lfloor \log_{2}n \rfloor - 1$ & $120n^{3}-1.67*2^{22}$ \\ \hline
     ECC & Banegas \cite{banegas2021} & $7n+\lfloor \log_{2}n \rfloor + 9$ & $48n^{3}+8n^{\log_{2}(3)+1} + n^{2}(352\log_{2}n + 512)$\\ \hline
    \end{tabular}
    \caption{\textbf{State-of-the-art resource estimates for quantum computers to break public-key cryptosystems.} This table shows the resource estimates for fault-tolerant quantum algorithms used to attack the RSA and ECC cryptosystems. $q$ refers to the circuit width, not the number of physical qubits required to run such circuits. The depth $d$ gives the number of algorithmic time-steps required, and does not refer to the amount of actual (``wall clock") time required.}
    \label{tab:q-cryptanalysis}
\end{table}

 We emphasize that in these resource estimates, $q$ refers to the number of \emph{logical} qubits required by the circuit, and \emph{does not} refer to the number of \emph{actual} or physical qubits a quantum computer would need to realize running that circuit. Logical qubits must be \emph{distilled} using quantum error correction (recall the end of Section \ref{sec:backdrop}). This distillation process introduces additional overhead with respect to both the number of physical qubits required, as well as the runtime of the algorithm, which is \textit{not} accounted for in the notion of circuit depth $d$ used here. And here, $d$ provides a measure of the total program complexity (in terms of primitive timesteps required), \textit{not} the total (physical) runtime of the algorithm. 

Also as an example of how the required resources will generally be greater than the ``bare" numbers, it should be noted that \cite{gidney2021} provides a detailed analysis of factoring RSA-2048 with a characteristic underlying gate error rate of $10^{-3}$ using a surface code and an efficient implementation of Shor’s algorithm. With this error rate, the \textit{final} number of logical qubits was estimated to be $q=14238$: more than a factor of 2 larger than plugging $n=2048$ into the equation for $q$ in Table 1 (namely, $q=6190$). The reason is that additional auxiliary qubits are needed to implement gate operations and for data movement.

The input to these resource estimates is the key length $n$. Although it may be intuitive to the reader that increasing $n$ increases the protection of the information encrypted using a given cryptosystem, we make explicit the relationship between $n$ and the number of bits of security the key provides. This relationship is a necessary ingredient in what follows. Thankfully, relating $n$ to number of bits of security is known!

For an $n$-bit RSA key, the approximate number of bits of security, $x_{\mathrm{RSA}}$, is \cite[Formula 1]{FIPS140-2ig}
\begin{equation}
\label{eq:rsa-bits}
    x_{\mathrm{RSA}}(n)= \frac{1}{\ln 2}\left(\sqrt[3]{64/9}*\sqrt[3]{(n\ln 2)}*\sqrt[3]{[\ln(n\ln 2)]^{2}} - 4.69 \right).
\end{equation}

For FFCDH, if the length of the public and private keys are $n$ and $m$ respectively, then the corresponding number of bits of security is
\begin{equation}
\label{eq:ffcdh-bits}
    x_{\mathrm{FFCDH}}(n,m)= \min (x_{\mathrm{RSA}}(n), m/2).
\end{equation}
Hence, as long as $m > 2x_{\mathrm{RSA}(n)}$, the security is limited by that of the public key, and is equivalent to the security of an RSA key of length $n$. We assume this in the analysis below.

For ECC, as mentioned previously, if the modulus of the prime number used to generate finite field for the curve is $n$ bits, then the number of bits of security is half that:
\begin{equation}
\label{eq:ecc-bits}
    x_{\mathrm{ECC}}(n)= n/2.
\end{equation}

With Table \ref{tab:q-cryptanalysis} and Equations \eqref{eq:rsa-bits}, \eqref{eq:ffcdh-bits}, and \eqref{eq:ecc-bits} in hand, we are now in a position to reason about how increasing the number of bits of security changes both the required quantum compute resources, and the upper bound on the required logical error rate. 

\begin{figure}
    \centering
    \includegraphics[width=\textwidth]{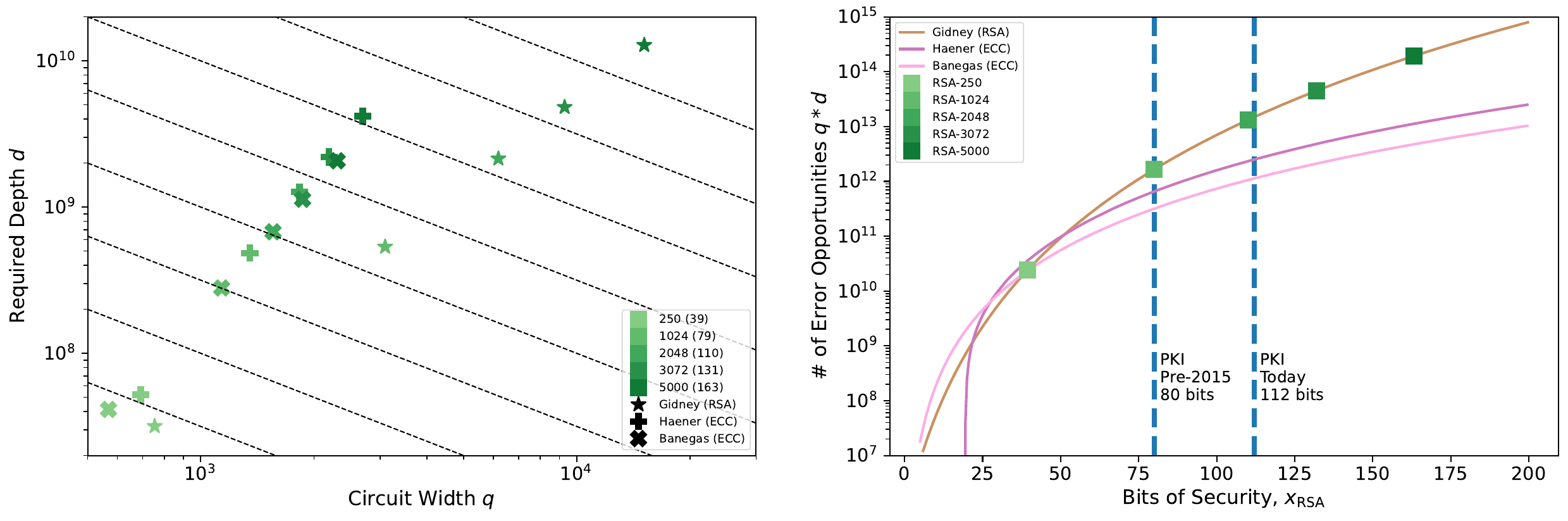}
\caption{\textbf{Quantum Algorithms for Cryptanalysis.}  \textbf{Left:} Circuit-model-based resource estimates for RSA and ECC. For a fixed number of bits of security, we plot $q$ and $d$ from Table \ref{tab:q-cryptanalysis} based on current implementations of Shor's algorithm. Shading gets darker with increasing number of bits of security (specified as ``RSA key size (bits of security)" in the legend). Dashed lines indicate constant values of the product $q*d$. \textbf{Right:} Error Opportunities ($q*d$)  for RSA and ECC as a function of bits of security, $x$. As the number of bits of security increases, the number of error opportunities for quantum algorithms attacking the RSA cryptosystem grows more rapidly than those for attacking ECC.}
\label{fig:resources}
\end{figure}

In the left side of Figure \ref{fig:resources}, we sweep across varying key lengths for RSA (fixing the size of the key length for ECC to yield the same number of bits of security) and trace out the circuit width $q$ and depth $d$ required to break them with current implementations of Shor's algorithm. The figure shows that, as the number of bits of security increases, the resources required for breaking RSA (both $q$ and $d$) increase more quickly than for ECC. This implies that for a given amount of security, data encrypted using RSA will be generally harder to decrypt using quantum computers than data encrypted using ECC. Recent optimizations for factoring using Shor's algorithm \cite{regev2023efficient,ragavan2023optimizing} will likely decrease the scaling of $d$ for RSA, but not decrease $q$.

In the right side of Figure \ref{fig:resources}, we plot the number of error opportunities $(q*d)$ as a function of the number of bits of security for current implementations of Shor's algorithm. Increasing the cryptographic key length $n$ increases both the number of bits of security, as well as the quantum compute resources required to break it. (Recall Table \ref{tab:q-cryptanalysis}: both $q$ and $d$ increase with $n$.) Hence, \textit{the number of error opportunities increases with the number of bits of security for the key}. This is clearly seen in the figure. A few specific RSA key sizes are labeled for the reader.

While the number of error opportunities grows with the number of bits of security, the rate of growth depends on the cryptosystem. The right side of Figure \ref{fig:resources} shows that for a fixed number of bits of security, the number of error opportunities is much lower for ECC than RSA. (This follows from the previously-observed fact that $q$ and $d$ for ECC are less than RSA.) \textbf{\textit{This implies that, given current resource estimates, ECC is more vulnerable to a successful quantum-based attack: the tolerable error rates for ECC are much higher for a successful quantum-based attack.}} For example, RSA-1024 is five times more difficult to crack as 160-bit or 163-bit ECC (as measured by the number of error opportunities).
We note however, that this assessment may change in the future with algorithmic improvement, such as recent developments for factoring that have not been fully assessed  yet \cite{regev2023efficient,ragavan2023optimizing}.

Also indicated in the right side of Figure \ref{fig:resources} is the evolution of the recommended standards for public-key cryptography. By increasing the number of bits of security from 80 (pre-2015 recommendation) to 128 (today), the number of error opportunities for both RSA and ECC go up by almost an order of magnitude. In turn, this means that by migrating from the old recommendation to the new one, the information stored with those keys is more secure against attacks by quantum computers. It should be noted that the largest RSA key factored by purely-classical means is RSA-250 \cite{rsa-250-factoring}, which is labeled in the figure.

Recall one of the observations made for Figure \ref{fig:circuit-resource-estimate}; namely, it is expected near-future quantum computers should be able to run circuits with a width of $\sim 1000$ and between $\sim 1K$ and $\sim 10K$ $T$ gates. Assuming that a comparable number of non-$T$ gates can be run as well, then the largest-possible sized circuit which could be run reliably has a size on the order of $10^{4}$. (It's not $10^{7} = 1000*10^{4}$ because the number of $T$ gates is the \emph{total} number, not the number per qubit.) 
Looking at the right side of Figure \ref{fig:resources}, being able to run circuits of such a size does not pose a threat to cryptography.

\begin{table}
    \centering
    \begin{tabular}{|c|c|c|c|c|c|c|}
    \hline
        \textbf{Year} & \textbf{$N$} & \textbf{$n =\lceil \log_{2}(N)\rceil$} & $x_\mathrm{RSA}(n)$ & \textbf{Factors} &  \textbf{Notes} \\ \hline
        2001 \cite{vandersypen2001} & 15 & 4 &N/A & 3, 5    & NMR System\\ \hline
        2011 \cite{Martin_Lopez_2012} & 21 & 5 & N/A & 3, 7    & Photonic System \\ \hline
        2019 \cite{Amico_2019}  & 35 & 6 &N/A & 7, 5  & Superconducting System  \\ \hline
        2019 \cite{Dang2019optimisingmatrix} & 961307 & 20 & 5.9 & 619, 1553   &  Simulator \\ \hline
        2023 \cite[Table III]{lee2023optimizing} & 247 & 8 &0.26 & 13,19    & Simulator \\ \hline
        2023 \cite{Willsch_2023} & 549755813701 & 39 &11.7 & 712321, 771781   & Simulator; iterative algorithm \\ \hline
    \end{tabular}
    \caption{\textbf{State-of-the-art, non-trivial implementations of Shor's algorithm for factoring integers (by year).}  Here, we focus on works which use \textit{non-trivial} implementations of the algorithm, in which a minimal (or no) amount of prior information about the solution is used to implement or compile the algorithm prior to running it on hardware or a simulator. For some integers, the number of bits of security they would provide if they were used as keys in an RSA cryptosystem is negative, indicated as N/A.}
    \label{tab:shor-hero}
\end{table}
Looking towards future quantum computers, estimating how much time remains before RSA-2048 keys can be broken is quite difficult. To date, neither experiments nor simulations thereof have tackled the problem of factoring integers at scales close to the aforementioned 829-bit RSA-250 number, much less RSA-2048 keys. Table \ref{tab:shor-hero} tabulates state-of-the-art implementations of Shor's algorithm (either on real quantum computing hardware, or simulators). We note that some implementations leverage compilation techniques and favorable parameter choices for the algorithm (as described in \cite{Smolin_2013}) which make the implementation trivial in nature. The results in Table \ref{tab:shor-hero} are non-trivial implementations, and thus provide an accurate representation of where the field currently stands in terms of using Shor's algorithm to factor integers. It is clear that both simulation and hardware experiments do not come close to realizing credible attacks on cryptographically-relevant RSA key sizes. In light of the prior paragraph, this is understandable: the hardware simply cannot run large enough circuits to tackle cryptography.

Many analyses have been done to assess the risk quantum computers pose to cybersecurity. One suggests RSA-2048 is unlikely to be factored ($<5\%$ confidence)  before 2039, based on a  statistical model extrapolating current quantum computing hardware progress \cite{sevilla2020}. Another, via a survey of quantum computing experts for the ``2022 Quantum Threat Timeline Report”, gives a 50\% probability for the same problem by 2037 \cite{mosca2022}. The 2023 version of the self-same report, which comparing prior estimates for the likelihood RSA-2048 could be broken within the next 5 years, shows expert assessments continue to assign this occurrence as being ``Extremely Unlikely ($<1\%$ chance)" \cite[Figure 10]{mosca2023}. An analysis of the energetic requirements needed to operate a cryptographically-relevant quantum computer (CRQC) was performed in \cite{parker2023estimating}, concluding ``...Even if a CRQC is eventually built, merely operating it would probably remain the domain of nation-states and large organizations for a significant period of time". Reference \cite{gidney2021} gives a physical resource estimate for using Shor's algorithm on RSA-2048 integers  of 20 million physical qubits and 8 hours of runtime\footnote{Note that many assumptions go into deriving those numbers: changing them would generally change the required number of physical qubits and/or runtime.}.

All of these analyses suggest that near-future quantum computers will not pose a risk to cybersecurity. However, given the significance of the potential threat, transitioning to new cryptographic primitives is necessary. This significance is highlighted in part by the USG's National Security Memorandum 10 (NSM-10) \cite{nsm-10} of May 4, 2022 entitled \textit{National Security Memorandum on Promoting United States Leadership in Quantum Computing While Mitigating Risks to Vulnerable Cryptographic Systems}. This memorandum directs USG to mitigate ``as much of the quantum risk as is feasible" by 2035. In addition, the NSA's Commercial National Security
Algorithm (CNSA) Suite 2.0 \cite{CNSA2.0} directs that by 2030, certain components of national security systems will need to use the algorithms documented in CNSA 2.0.

In the next section, we briefly discuss how cryptography has been evolving in light of the threat posed by quantum computers, and how organizations can prepare now to transition their digital systems to such cryptosystems.

\section{Evolving Cryptography for the Quantum Era}
\label{sec:quantum-safe}
\addtocontents{toc}{%
  \smallskip\protect\parbox[t]{\textwidth}{\textit{Discusses how cryptographic primitives and infrastructure will need to be updated to address the threat of quantum computing, and how organizations can begin getting quantum safe.}}\par~\\}

Given the threat of quantum computing to cryptography, identifying, standardizing, and implementing new cryptographic systems which are believed to be resistant to attacks from \textit{both} classical and quantum computers is one of the most pressing tasks in the cybersecurity community today. We briefly review the three primary lines of work in doing so: quantum key distribution (QKD), quantum random number generation (QRNG), and ``post-quantum" cryptography\footnote{These techniques are often captured under the common umbrella of ``quantum-safe cryptography'', since they are designed to also be safe against quantum-enabled attacks. In some of the literature the term quantum-safe cryptography is equated with post-quantum cryptography only. In this article we take the broader interpretation.}. In doing so, we emphasize the importance of considering the evolution of cryptography in light of quantum computers as being about establishing ``defense-in-depth" through a judicious choice of relevant cryptographic methods. These methods must consider the different ways that cryptography is used to protect computing, for example, to protect the confidentiality of data when it is stored or transmitted, to authenticate entities such as people and software updates, and to provide repudiation in the case of digital signatures.  

The reader may be wondering what urgency, if any, exists for evolving the use of cryptography. In light of the analysis in the previous discussions, it should be clear that near-term quantum computers should not be able to crack currently-used public-key infrastructure, especially if the keys used conform to the updated recommendations from NIST (namely, to use at least 112 bits of security). So why worry about addressing what seems, at first glance, to be a non-existent threat to currently-used cryptosystems?

The urgency depends not only on the estimated time when adversaries will be able to deploy quantum-enabled attacks, but also the shelf-life of information and the time to migrate to quantum-safe cryptography \cite{mosca2018cybersecurity}.  In brief, urgent migration is required in cases where data is sensitive and has a long shelf life, and in cases where systems are complex with many dependencies and migrations are complex. First, while the resources required to break current cryptosystems are large and not expected to be available in the short term, advances in the understanding of quantum algorithms and of fault-tolerant quantum error correction have \textit{reduced} them over time. Lacking any \textit{a priori} reasons why this trend cannot continue, it is reasonable to assume future advances will continue to reduce the resources required. Phrased another way, the threat may arrive sooner than expected, and the likelihood of quantum-enabled attacks in the next few years may already exceed the risk tolerance of many organizations. Second, data may actually be at risk already, even though capable-enough quantum computers do not yet exist.  This risk is referred to as \emph{harvest now - decrypt later}.  The idea is an adversary could acquire encrypted data, and simply hold onto it despite being unable to decrypt it.  For data which is sensitive and needs to be kept secure for a long term, this risk necessitates evolving the cryptography used to secure it\footnote{Far from being an idle point of speculation, it is generally acknowledged this tactic is currently being carried out today \cite{qtech-hndl}. In addition, in the context of blockchain or other decentralized financial ledger technologies (DeFi), adversaries do not need to harvest the data -- it is already exposed publicly on the blockchain itself \cite{pld-defi}.}.

In other words, data encrypted using current methods may actually be already collected by an adversary who is simply waiting for the development of a capable-enough quantum computer to decrypt it. Third, updating the cryptography used throughout an organization can be a time-consuming process. Based on the past history of cryptographic transitions, this process will take at least a decade (likely longer) and needs to be completed before quantum computers advance to the point they break current cryptosystems.   Most organizations are not naturally suited to undertaking such a task, and will require substantial time and help to do so. The best time to lay the foundations for getting started is now.

\subsection{Quantum Safe Approaches}
\label{sec:q-safe-approaches}
In short, even though near-term quantum computers won't be able to attack current cryptosystems, prudence necessitates grappling with the fact cryptography will need to evolve for the quantum era.  Below we briefly describe each of the approaches currently being pursued: quantum key distribution (QKD), quantum random number generation (QRNG), and post-quantum cryptography (PQC).
As we will discuss later in this sub-section, a strategy for getting quantum safe must incorporate, at minimum, the use of PQC, and the other methods may be used to further enhance security and protection.

\subsubsection{Quantum key distribution (QKD)}

QKD enables the exchange of symmetric keys through an authenticated public channel without computational assumptions.  The comparable classical methods for achieving key exchange require a non-trivial computational assumption equivalent to a secure trapdoor predicate. QKD involves preparing and sending quantum states from one party to another. These states are measured by both parties, and the relationships between the measurement outcomes are used to generate a shared random key. 
Ever since the method was first described \cite{bennett2014quantum}, much work has been done in the academic and commercial space for realizing deployments \cite{idquantique-qkd,kets-qkd,quintessencelabs}. It is currently deployed commercially in point-to-point and trusted repeater networks, with plans to deploy long distance QKD services via satellite systems and/or with quantum repeaters once they are available. 

Some nation-states -- notably, China -- have pursued varied and extensive QKD deployments \cite{liao2017satellite,liao2018satellite,chen2021integrated,chen2021twin,chen2021implementation}.  In the US, the Department of Energy has engaged in research using QKD for the purposes of securing the electrical grid \cite{9797980,9405393,alshowkan2022authentication,8908470,williams2021implementation}. Activities are underway regarding the standardization of these methods \cite{stanley2022recent}. For example, the mission of the OpenQKD project in the EU ``is the establishment of QKD-based secure communication as a well-accepted, robust and reliable technology instrumental for securing traditional industries and vertical application sectors, and to prepare the deployment of a Europe-wide QKD-based infrastructure in future." \cite{openqkd}. Also, the European Telecommunications Standards Institute (ETSI) has been working to create and promote QKD standards \cite{etsi-qkd}.  Other standardization efforts are underway across the EU as well \cite{van_Deventer_2022}.

The journey to large-scale production deployments of QKD networks will need to address known issues, including establishing trusted nodes for QKD deployments \cite{ncsc-guidance,nsa-qkd,ANSSI-QDK-guidance,etsi-qkd-challenges}.  It may be possible that future versions of QKD (especially ``QKD 2.0": device-independent QKD using entanglement distribution over networks) can resolve some of these key issues in securing QKD systems \cite{renner2023debate}. As a result of current limitations, the usage of QKD is not presently recommended for national security systems in the US or UK \cite{ncsc-guidance,nsa-qkd}. 

\subsubsection{Quantum random number generation (QRNG)}

QRNG is a method to produce random numbers based on measuring a quantum-mechanical system. Several private firms have created products (namely, devices) to offer QRNG capabilities \cite{quantum-origin,idquantique-qrng,kets-qrng,qrypt,quintessencelabs}. The random numbers generated by QRNG devices can then be fed into cryptographic algorithms to generate keys. This approach has been touted as a way to generate ``truly random" randomness and keys \cite{evolutionq-qrng}.

Similar to QKD, there are challenges for QRNGs such as the possibility of implementation attacks \cite{ncsc-guidance,ma2016quantum}. In the US, NIST has developed certifications for randomness which enable end-users of QRNGs to have high confidence in the randomness of the bits they output \cite{nist-rng}. And in the EU, the Federal Office for Information Security in Germany has also put forward specifications and certifications for RNGs \cite{bsi-rng-guidance}.

\subsubsection{Post-quantum cryptography (PQC)} 

PQC is a set of purely-classical means for encrypting and decrypting data using algorithms which are presumed to be hard for both quantum \textit{and} classical computers. Post-quantum cryptosystems rely on different mathematical problems, such as lattice-, code-, or hash-based problems which are believed to be difficult for quantum computers to solve \cite{nist-pqc-overview,kumar2020post,buchmann2016post}. 

There are concerted global efforts facilitating the development and deployment of post-quantum cryptographic systems \cite{nsa-qsc,nist-pqc}. NIST announced in February 2016 it would begin standardization efforts for post-quantum cryptography, by working with the international community and running a competition-like process to find and eventually select new algorithms to replace the existing public-key algorithms which are vulnerable to quantum computers \cite{NISTPQCannouncement}. As the past 7 years has shown, the process of designing new quantum-resistant algorithms, evaluating their security, and standardizing them is not a short one!

After NIST's initial call, 82 total proposals were submitted.  These submissions were evaluated based on their security, performance, and other characteristics during a series of three rounds, with the most promising schemes moving on to the next round of evaluation. The $3^{\text{rd}}$ round consisted of 7 finalist algorithms, and 8 alternate algorithms.

In July of 2022, NIST announced the selection of four PQC algorithms for standardization \cite{NISTPQCselection}. The draft standards are currently available for public comment and are expected to be published in early 2024 \cite{Moody_future,NISTPQCStandardization1,NISTPQCStandardization2,NISTPQCStandardization3}. An additional four algorithms were also moved to a $4^{\text{th}}$ round of evaluation for consideration for future standardization \cite{NISTPQCround4}.  Simultaneously, NIST issued a new call for quantum-resistant digital signatures to complement the ones it had already selected \cite{NISTPQCsignatures}.  The evaluation process for these signatures will last several years before any of them could be selected for standardization.  As a result of the selection of a first set of algorithms for standardization, NSA also announced the CNSA 2.0 suite \cite{CNSA2.0}, together with a FAQ \cite{CNSAFAQ}, which builds on the successful NIST process and includes algorithms selected by NIST.

Similar efforts are underway also outside the U.S. in the EU \cite{EUPQC}, China \cite{ChinaPQC}, and with international standards bodies, including ISO \cite{ISOPQC} and the IETF \cite{IETFPQC}. 

\subsection{Laying the Foundations for Getting Quantum Safe}
\label{sec:q-safe-foundations}
While most organizations rely on publicly-scrutinized and standardized algorithms and protocols, this does not mean they can simply wait until standards are available. Even once a standardized algorithm is available, it can take many years to deploy it in a production system.  To shorten their timeline to benefiting from the protection of a new cryptographic algorithm, organizations can perform much of the testing and integration work in parallel with the standardization process.  For example, strong open-source implementations of quantum safe algorithms have been available for many years to enable organizations to test their impact in the most important applications \cite{open-quantum-safe-project}.  Many of the largest global IT vendors have been doing so in some of their products for several years, and have even offered quantum safe alternatives for some of their products \cite{ibm-z-16-1,ibm-z-16-2,ibm-quantum-safe-tape,aws-pqc-1,aws-pqc-2,aws-pqc-3}.

Furthermore, organizations don't have to choose between using a method which is known to be compromised by quantum computers and using a not-yet-standardized key exchange method.  There are standardized mechanisms for combining keys exchanged with quantum safe algorithms (or other means) with today's standardized algorithms without losing certifications associated with today's standards. See, for example, a technique outlined by NIST in its guidance on how hybrid key-establishment can be performed in a FIPS 140-approved mode of operation \cite{nist-transition-faqs}.  While there are clear advantages to such a hybrid approach, there are also some potential disadvantages such as increased complexity and interoperability concerns.  Both Germany and France will require hybrid solutions for their national security systems, while the US will not \cite{ANSSIhybrid, BSIhybrid, CNSAFAQ}.

It should be noted there are other well-established means of exchanging cryptographic keys, such as through the use of pre-shared keys and methods relying on symmetric key algorithms and trusted third parties. For example, the NSA's guidance on Commercial Solutions for Classified (CSfC) specifies that ``Symmetric Pre-Shared Keys (PSKs) should be used instead of or in addition to asymmetric public/private key pairs to provide quantum resistant cryptographic protection of classified information within CSfC solutions." \cite{nsa-cfsc}. These methods are not intended to be a general purpose replacement for public-key cryptography, but serve a critical role in providing defense-in-depth and protecting systems where the possibility of breaking an asymmetric key exchange is deemed too risky.

Thus, organizations have a well-established path for mitigating against harvest now - decrypt later attacks today, and beginning their cryptographic evolution while standards are completed. This proactive approach also has the benefit of engaging the providers of cryptographic solutions earlier in the migration process and enabling a higher level of product maturity and reliability sooner, and the development of a trustworthy supply chain. Therefore, even before standards are finalized, these new cryptographic approaches can be securely integrated alongside standardized classical protocols.

While QKD and other alternatives to asymmetric algorithms can address specific scenarios (such as the use of Pre-Shared Keys in high assurance encryptors) and more generally offer opportunities for greater defense-in-depth, the layer of defense currently provided by asymmetric algorithms is fundamental to most widely deployed digital platforms and organizations worldwide are focused on realizing a readily-available and essentially ``drop-in" replacement for their existing cryptography.  Of course, we cannot ignore the possibility  the new post-quantum algorithms may be broken at some point by classical or quantum algorithms. And one can't assume that next time we'll also have several decades between discovering the algorithmic break and adversaries implementing attacks based on them. Therefore it is important for the resilience of digital systems to keep researching, developing, and standardizing strong alternatives, as well as designing systems with the agility to change algorithms relatively quickly and easily. Furthermore, especially given the increasing reliance on cryptography, in addition to greater agility, a defense-in-depth approach is prudent, especially for more critical systems.

\subsection{Practical Considerations for Getting Quantum Safe}
\label{sec:practical-q-safe}
Every organization depends on digital infrastructure communications – all of which inherently rely on cryptography to ensure trust. In this sense, organizations consider cryptography as their ultimate line of defense. Both trust and protection are  at risk as asymmetric cryptography is likely to be broken when a cryptographically-relevant quantum computer becomes available and due to the ``harvest now -- decrypt later" attack discussed earlier. Hence, it is necessary for organizations to undergo an evolution in their use of cryptography, one which involves the deployment of post-quantum cryptography within their operations.

The IT landscape for many organizations has become increasingly complex over the years, encompassing many facets: digital applications developed internally, IT capabilities inherited through acquisitions, third party software products, services from cloud providers, consumption of SaaS capabilities, etc. Hardware itself is changing, with devices which provide critical services such as point of sale terminal in a retail setting, medical data telemetry, or an IoT device deployed in manufacturing or mining scenarios also relying on establishing trust using cryptography. The scope of what needs to be upgraded or remediated to be quantum safe is very vast. The transition will take years of work for an enterprise as it migrates to quantum-resistant algorithms. This journey must be well-planned and incrementally executed.

Organizations which seek to undergo this transition will find their transformation journey to become quantum safe is quite achievable. Several resources exist for organizations to get started on their journeys, including offerings provided by the private sector \cite{ibm-quantum-safe,evolutionq,sandbox-quantum-safe,pqshield,quantum-exchange}, and guidance from NIST's National Cybersecurity Center of Excellence \cite{nist-nccoe}, the US government \cite{usg-pqc}, the UK's NCSC \cite{ncsc-guidance2}, the Canadian Forum for Digital Infrastructure Resilience (CFDIR) \cite{cfdir-q-ready}, and the Financial Services Information Sharing and Analysis Center (FS-ISAC) \cite{fs-isac-q-ready}. Finally, Section 4.4 of \cite{NAP25196} discusses a framework security experts and policy makers can use to assess the difficulties in making the transition to quantum-safe cryptography. The results for organizations which take this journey in earnest will include cost-efficient remediation, improved cryptographic posture, enhanced compliance, and reduced regulatory risk and benefit longer-term through a more agile cybersecurity management process. 

Drawing on the experience of the authors and the above resources, some important steps that should be undertaken for this transformation include:

\begin{enumerate}
    
\item \textbf{Inventory cryptography usage.} The first step is to ensure an organization has a very comprehensive view of its usage of cryptography. This is achieved by inventorying both the static \textit{and} dynamic use of cryptography. Scanning of cryptography usage in custom applications used within an organization provides a static view, and scanning the organization’s network for cryptography calls provides a dynamic view of cryptography usage. When combined, these methods provide a comprehensive inventory of cryptography usage across an entire organization.  This inventory of cryptography usage is best represented in a standard form, such as a cryptographic bill of materials (CBOM).

\item \textbf{Identify and prioritize crown jewels.} A comprehensive inventory of cryptographic usage is a necessary starting point. However, more information is needed from a planning perspective to create actionable insight. It is critical that the most critical data and assets of the organization, (i.e., crown jewels) are identified and safeguarded first. This is achieved by classifying the various data assets within the organization and using these as filters to prioritize and plan. This classification and selection must consider both technical and business perspectives. This prioritized view will then be used to develop a transformation plan or roadmap. 

\item \textbf{Develop a transformation roadmap.} It is not possible, or pragmatic, to transform every application and system within an organization in one massive initiative. Transformation efforts must be broken into several phases with each phase building on the collective experiences and capabilities of the previous phase. Developing an incremental transformation roadmap is a very important step and must be done with great thought and depth of understanding of the challenges involved. Typically, transformation projects should start small, learn from each iteration, build reusable components, and eventually establish comprehensive approaches to be executed diligently over time. 

The scope of the transformation must include systems within the control of the organization as well as ensuring that appropriate external or third-party entities that the organization integrates also complete their transformation. This software supply chain view and the respective CBOM is a great way to identify and map the dependencies and diligently use apply this to chart the transformation roadmap.

While the transformation roadmap will likely be centered around migration of the existing algorithms to post-quantum algorithms, this is also the time to articulate the plan in case a system centred around post-quantum algorithms is broken (whether the break is relatively easily fixable or a more profound and systemic break). Other layers of defense such as those leveraging pre-shared keys, symmetric cryptography and quantum cryptography can be explored and tested.

\item \textbf{Begin the transformation to become quantum safe.} Once an actionable transformation roadmap has been developed the process of execution can begin. As indicated earlier, the transformation journey will typically take several years. Given this duration, it is important that the transformation roadmap be periodically validated with the learnings from ongoing transformation efforts. Environmental considerations also need to be factored into this process. This constant validation and fine tuning of the transformation plan will keep it current, valid, and relevant.

One key aspect of this transformation step is ensuring that \textit{crypto-agility} is part of the transformation. Ensuring that future cryptography related changes do not require core changes to the application is what is typically referred to as crypto agility. This enables ease of transition from one post quantum cryptography algorithm to another in the future – should the need arise. Therefore, activities must include enabling crypto agility as part of the transformation journey. Another key aspect of the transformation step is to not assume that the effort is one of simply replacing old cryptography with the newly standardized post quantum cryptographic algorithms. While this is indeed a vital component – it is not necessarily the most efficient or effective method. Identifying pre-defined remediation approaches or pattern, leveraging modern deployment practices for transforming an application or system is important. It is very likely that organizations with a lot of applications typically can establish and use multiple new patterns of cryptographic deployment repeatedly across their IT landscape. 

\item \textbf{Gain familiarity and expertise in the use of new algorithms and tools.} Establishing a credible level of competency in the deployment and operational use of quantum-safe cryptography, executed in parallel with the previously enumerated steps, is another critical step. The new post quantum cryptography algorithms are quite complex and quite different from existing algorithms. Understanding the appropriate usage of the algorithms, their performance profile and resource consumption characteristics is very important. Other quantum-safe alternatives will also require specialized expertise to properly deploy.

This is accomplished by establishing a Center of Excellence; identifying dedicated cryptographic and network engineers, security architects, application developers and project management personnel and providing them an environment and an opportunity to experiment and learn. It is extremely useful to have a ready-made operational environment that supports targeted pilot projects for testing new algorithms and assess their deployment with the organization operating environment. 

\end{enumerate}

\section{Conclusion and Discussion}
\label{sec:conclusions}
\addtocontents{toc}{%
  \smallskip\protect\parbox[t]{\textwidth}{\textit{Recaps the themes and ideas discussed in the main text, and suggests recommendations as to what organizations, academia, the commercial quantum computing sector, and policy makers/regulators should do.}}\par}

Quantum computing holds promise as a disruptive technology that extends the frontiers of computation. This promise could tackle and solve some of our most pressing grand challenges. At the same time, the realization of a cryptographically-relevant quantum computer would have serious implications for cybersecurity. This article provided a broad and thorough survey of the ways in which quantum computers may impact society and their ramifications for cybersecurity. We briefly review the major themes and ideas below, before suggesting some recommendations.

In Section \ref{sec:backdrop}, we gave a survey on the backdrop (political, cultural, and scientific) against which a nascent commercial quantum computing ecosystem has developed. From this, it is clear quantum computing has garnered tremendous interest from the private sector, with commercial companies spearheading major research initiatives to make quantum computing practical for end-users and their applications. The three major themes over the past 5 years within this ecosystem have been: (a) building quantum computers capable of tasks which are difficult for classical computers to solve, (b) leveraging quantum computers to explore problems of scientific and commercial interest, and (c) building on these advances towards the development of a fully fault-tolerant quantum computer. Each of these themes has required much investment in research, systems engineering, and ecosystem-building. Support from government initiatives around the world has supported this investment, which we hope to see continue and grow in the coming years.

In Section \ref{sec:benefits}, we took a comprehensive look at the literature for applications of quantum computational science to practical problems (with a particular emphasis on the commercial uses). Figure \ref{fig:circuit-resource-estimate} shows that although there are a wide variety of problems for which resource estimates have been done, there is still a lack of commercially-relevant ones whose circuits can be executed using current and near-future quantum computers. Identifying such problems is essential for catalyzing broad adoption of quantum computing. We discussed how circuit knitting, variational algorithms, and error mitigation may help unlock useful quantum computing in the near-term, because they can reduce the size of the circuits required. We surveyed the efforts private industry has launched to explore the potential of quantum computing to transform their businesses.  Based on what is currently known, it is the best guess of the authors that variational algorithms are critical to explore for near-term economic applications, as recent experiments and results are increasingly suggestive of the potential for near-term quantum computers to exhibit utility for problems of academic or scientific interest. In addition, simulating the time dynamics of quantum spin systems seems to be a quite promising area where quantum computers could make a meaningful impact in the near-term.

Section \ref{sec:national security} provided a history of the evolution of public-key cryptosystems, and an examination through the lens of logical resource estimates of the capabilities needed by a quantum computer in order to attack the public-key infrastructure currently in place. We proposed using a conservative upper bound on the required logical error rate as a framework for reasoning about the risks a given quantum computer would pose to a given cryptosystem and key size. Our analysis shows that increasing the number of bits of security used by a given cryptographic key increases the quantum compute resources needed to break it. (Namely, the maximal tolerable logical error rate goes down.) As noted in the section, advances in quantum algorithms would generally lessen the required capabilities needed to field a cryptographically-relevant quantum computer. Our analysis also shows that, to date, no known quantum computer exists with the required capabilities to attack public-key infrastructure, either at the pre-2015 recommended security strengths, nor at those recommended for today.

Finally, in Section \ref{sec:quantum-safe}, we discussed how, even in the absence of the development of a cryptographically-relevant quantum computer within the next several years, beginning the transition to quantum-safe cryptography is of paramount importance for ensuring security. Several approaches exist for creating such cryptosystems: quantum key distribution (QKD), quantum random number generation (QRNG), and purely-classical post-quantum cryptography (PQC). Focusing on PQC, we discussed the evolution of quantum-safe cryptographic standards, and shared ideas on how organizations can begin their transformation journeys. In particular, it is important for organizations to develop a comprehensive view of their usage of cryptography and the vulnerabilities they have -- a task enabled by tools being built today. Remediation of these vulnerabilities will need to proceed in a systematic and thoughtful way to ensure all places where vulnerabilities exist are remediated in a consistent and prioritized fashion.

In our view, a thoughtful approach to quantum computing necessitates engagement with both the potential applications of quantum computational science \textit{and} the transition to quantum-safe cryptography. To these ends end, we suggest that:
\begin{enumerate}
    \item Commercial providers of access to quantum computers should continue to...
    \begin{enumerate}
    \item ...invest in improving the quality of the underlying hardware. As the ``base", so to speak, of the quantum computing technology stack, reductions in noise and error rates benefit all higher levels, including the ability of end-users to explore larger-sized quantum circuits on actual quantum computers.
    \item ...engage with academic groups to develop additional techniques for circuit knitting, error suppression, error mitigation, and error correction. Doing so is essential for extracting useful and practical value from near-term quantum computers, and facilitating advances towards large-scale, fault-tolerant quantum computing.
    \item ...nucleate communities of interest to stimulate discussion and engagement with end-users of quantum computers to identify potentially-promising areas where using quantum computers could make an impact, and chart aggressive research agendas. Examples of such communities include ones in high-energy physics \cite{dimeglio2023quantum,PRXQuantum.4.027001,humble2022snowmass,humble2022snowmass-2,beck2023quantum,bauer2022quantum}, healthcare/life sciences \cite{basu2023towards,emani2021quantum}, operations research \cite{abbas2023quantum}, and materials science \cite{alexeev2023quantumcentric}. These communities should seek to develop quantum-ready software and platforms.
    \item ...provide clear and transparent forward-looking roadmaps highlighting the major milestones they seek to achieve as they continue to commercialize their quantum computing activities. Such roadmaps help de-risk preparatory activities from end-users and other interested parties, and help minimize disruptive surprises.
    \end{enumerate}
    \item Organizations should...
    \begin{enumerate}
        \item ...develop and deploy a strategy for taking advantage of quantum computers as they mature. Such a strategy will of course depend on the particular organization, its strategies around technology more generally, and its interests in use of advanced compute capabilities. This said, organizations should be intentional about how they wish to use this technology.
        \item ...designate personnel to stay up to date on the latest developments in the direction of cryptographically-relevant quantum computers. These considerations are also important in light of the observation that it remains possible a new quantum algorithm may be developed which requires fewer resources than those presented in Figure \ref{fig:resources}, or perhaps even new purely-classical ones.
        \item ...transition to quantum-safe cryptography as soon as possible, subject to the final standardization of quantum-safe algorithms.  Given the potential risks to public-key cryptography, it is essential to identify vulnerabilities, and be prepared to remediate them in a timely fashion.
    \end{enumerate}
    \item Academic researchers should...
    \begin{enumerate}
        \item ...first continue to do the foundational research which has led to the hardware innovations, new algorithms, new approaches to error correction, new architectural concepts, and other ground-breaking work which both trains the workforce and generates novel concepts which can be adopted and built on by industry.
        \item ...should engage with commercial providers of access to quantum computers to  examine obstacles to scaling their research problems to take advantage of the capabilities being fielded. Quantum computers are becoming useful tools for novel research  \cite{baumer2023efficient,chen2023realizing,shtanko2023uncovering,PhysRevResearch.5.013183,farrell2023scalable,Pelofske_2023,chowdhury2023enhancing}, and such engagement allows the research community to push the frontiers of their own research and to aid industry.
    \end{enumerate}
    \item Commercial providers of access to quantum computers -- along with academic groups -- should continue their advocacy and education efforts around quantum computing and quantum-safe cryptography. These efforts are essential for catalyzing a broad-based adoption of these technologies, and also provide the necessary context required by policy makers, regulators, and the like for making informed decisions around this technology.
\end{enumerate}

In sum, we believe this paper provides a realistic and unexaggerated view of the risks and benefits of quantum computing technology and its applications.  The view is aligned with National Security Memorandum 10 (NSM-10) \cite{nsm-10}:
\begin{quote}
    Quantum computers hold the potential to drive innovations across the American economy, from fields as diverse as materials science and pharmaceuticals to finance and energy.  While the full range of applications of quantum computers is still unknown, it is nevertheless clear that America’s continued technological and scientific leadership will depend, at least in part, on the Nation’s ability to maintain a competitive advantage in quantum computing and QIS [Quantum Information Science]. 
\end{quote}

In conclusion, we restate the key promise of this technology.  While the attainment of a cryptographically-relevant quantum computer in the near or medium future remains conceivable, an assessment of the near-term economic impact of quantum computing should take into account the potential for the realization of economically-viable quantum computers to precede those with cryptographic significance.

To date, \textit{no proven application} of a quantum computer for \textit{economically-impactful problems} has been identified which could be run on a quantum computer whose capabilities would not also enable it to attack cryptography. However, it is the belief of the authors (and many of our colleagues\footnote{For beliefs of others in the field, see Section 4.6 of \cite{mosca2023}, the caption of Figure 13 in particular, which we quote verbatim: ``We asked the experts to indicate the likelihood for commercial applications of ``early" quantum computers / quantum processors not powerful enough to be directly relevant from a cryptographic perspective. Not all experts
expressed an opinion in this sense, but among those who did, more than half indicated a likelihood of about 50\% or more within 5 years."}) that with:
\begin{enumerate}
\item anticipated improvements of quantum hardware and software,
\item potential improvements in quantum algorithms, 
\item advances in quantum error mitigation,
\item the continued, determined pursuit of real-world applications by the private sector,
\item the current understanding about the resources required for cryptographically-relevant quantum computation, and
\item the expected/mandated transition to quantum-safe cryptographic protocols,
\end{enumerate}
that \textbf{\textit{there is a credible expectation that quantum computers will be capable of performing computations which are economically-impactful before they will be capable of performing ones which are cryptographically-relevant}}. 

This belief and potential promise is based on the remarkable advances made on both the hardware and software by academia and industry. Continued investment is vital for realizing the full potential of this technology. Moreover, such investment would still be fruitful even if the most substantial returns are realized over a comparatively longer time horizon: the cumulative end-user know-how and technological development would still be remarkably valuable, and the technology itself would be considerably de-risked by that time.
~\\~\\
In brief, it is essential to continue researching, developing, deploying, and using quantum computers -- a technology which will transform computing in the $21^{\mathrm{st}}$ century -- and to rapidly and expeditiously implement quantum-safe cryptographic protocols.
~\\~\\
It is our hope this work provides a comprehensive and comprehensible survey of quantum computing, its possible applications and risks, and the ways in which this technology is poised to transform our societies and civilization.

\section{Acknowledgements}

TLS acknowledges: (1) support from his family while working on this manuscript, (2) feedback from Sergey Bravyi, Jerry Chow, Antonio C\'{o}rcoles, Oliver Dial, Bryce Fuller, Dmitri Maslov, Patrick Rall, Omar Shehab, Sarah Sheldon, Matthias Steffen, Kristan Temme, and Ted Yoder, (3) informative discussions with Joseph Broz, Michael Beverland, and Pierre-Luc Dallaire-Demers, and (4) contributions from Jeffrey Brown and Ray Harishankar. In addition, TLS would like to acknowledge the entire quantum computing community for their efforts and works: without the total effort of this community, across all aspects of quantum information science, this paper would not have been possible.

CJW acknowledges helpful discussion with numerous individuals and parties from AWS, Google, IBM, Infleqtion, and Quantinuum. Special acknowledgements go to Chris Langer from Quantinuum and Robert Sutor from Infleqtion for providing useful perspectives relevant to the analysis and summary presented. The views and analysis presented in this work are solely of the authors.

WJZ acknowledges discussion and help from Christophe Jurczak, Olivier Tonneau, and members of the Quantonation and Unitary Fund communities.

MT acknowledges helpful discussions with Hasan Ali, Brad Lackey, Torsten Hoefler, Martin Roetteler, Mathias Soeken, Andrew Daley and Huang Hao Low.

The authors thank John Preskill for insightful and thoughtful comments on earlier versions of this manuscript. In addition, the authors thank Corey Stambaugh for feedback and comments.

Research at IQC is supported in part by the Government of Canada through Innovation, Science and Economic Development Canada (ISED).Research at Perimeter Institute is supported in part by the Government of Canada through ISED and by the Province of Ontario through the Ministry of Colleges and Universities.
~\\~\\
Disclaimer 1: The views expressed in this work are those of the authors only, and do not reflect those of IBM or Microsoft Corporation.
~\\~\\
Disclaimer 2: Any mention of commercial products or reference to commercial organizations is for information only; it does not imply recommendation or endorsement by NIST nor does it imply that the products mentioned are necessarily the best available for the purpose.
~\\~\\
Disclaimer 3: The opinions, recommendations, findings, and conclusions in this publication do not necessarily reflect the views or policies of NIST or the United States Government.

\newpage

\appendix

\section{Logical Resource Estimates}
\label{sec:circuit-resource-table}
\addtocontents{toc}{%
  \smallskip\protect\parbox[t]{\textwidth}{\textit{Provides a table of logical resource estimates used in Figure \ref{fig:circuit-resource-estimate}, along with the sources from which those estimates are drawn.}}\par}

In this appendix, we put a detailed table of the references and numbers used to populate Figure \ref{fig:circuit-resource-estimate}.

\begin{longtable}{|p{.75cm}|p{2cm}|p{1.5cm}|p{1.6cm}|c|c|c|p{1.7cm}|p{2.1cm}|} 
    \hline 
       \textbf{Row} & \textbf{Application}  &\textbf{Sub-Area} & \textbf{Problem} & \textbf{Qubits} & \textbf{$T$} & \textbf{Toffoli} & \textbf{Reference} &  \textbf{Notes} \\ \hline \hline
      1 &   Simulating Nature & CMP &$10\times 10$ FH & 236  & $7.1\times 10^{8}$ &  & \cite[Table IV]{Babbush_2018} & \\ \hline
      2 &   Simulating Nature & CMP &$10\times 10$ FH & 204  & $9.2\times 10^{7}$ & $1.3\times 10^{6}$ & \cite[Table I]{Kivlichan_2020} & \\ \hline
      3&  Simulating Nature&CMP & $2D$ TFIM  & 230 & $2.4\times 10^{6}$  &  &\cite[Table I]{beverland2022assessing} & \\ \hline
      4&  Simulating Nature & CMP & SS  & 131 & $6.1\times 10^{7}$ & & \cite[Table I]{Nam_2019} & Repeat-until-success\\ \hline
     5 & Simulating Nature & CMP & SS & 131 & $1.3\times 10^{7}$ & & \cite[Table I]{Nam_2019} &~"~ \\ \hline
     6 &    Simulating Nature & CMP &SS & 101 & $7.2\times 10^{6}$ & & \cite[Table I]{Nam_2019} & ~"~\\ \hline
    7 & Simulating Nature & CMP & SS & 100 & $1\times 10^{12}$ && \cite[Figs I \& II]{childs2018toward} & $N=100$; product formula \\ \hline
    8 & Simulating Nature & CMP & SS & 120 & $2\times 10^{10}$ && \cite[Figs I \& II]{childs2018toward} & $N=100$; quantum signal processing \\ \hline 
    9&    Simulating Nature & CMP & SS & 250 & $2\times 10^{11}$ && \cite[Figs I \& II]{childs2018toward} & $N=100$; Taylor series\\ \hline 
    10 & Simulating Nature & HEP & EFT & 6000 & $4\times 10^{12}$ & & \cite[Table 8]{watson2023quantum} & VC encoding \\ \hline
    11 & Simulating Nature & HEP & EFT & 10000 & $4\times 10^{12}$ & & \cite[Table 8]{watson2023quantum} & Compact encoding \\ \hline
    12 & Simulating Nature & Chemistry & FeMoCo & 135 & $3.5\times 10^{15}$ & & \cite[Table I]{reiher2017elucidating} & Quantitative accuracy \\   \hline
    13 & Simulating Nature & Chemistry & FeMoCo & 135 & $3.3\times 10^{14}$ & & \cite[Table I]{reiher2017elucidating} & Qualitative accuracy \\   \hline
    14 & Simulating Nature & Chemistry & FeMoCo & 378 & $2.1\times 10^{13}$ & & \cite[Table I]{Berry_2019} & ``Dirty ancilla" algorithm; RWSWT orbitals \\  \hline
    15 & Simulating Nature & Chemistry & FeMoCo & 437 & $2.0\times 10^{13}$ & & \cite[Table I]{Berry_2019} & ``Dirty ancilla" algorithm; LLDUC orbitals \\  \hline
    16 & Simulating Nature & Chemistry & FeMoCo & 3400 & & $1.3\times 10^{10}$ &  \cite[Table I]{von_Burg_2021} & Structure I, 52 orbitals \\  \hline
    17 & Simulating Nature & Chemistry & FeMoCo & 2142 & & $5.3\times 10^{9}$ & \cite[Table I]{Lee_2021} & RWSWT orbitals \\  \hline
    18 & Simulating Nature & Chemistry & FeMoCo & 2196 & & $3.2\times 10^{10}$ & \cite[Table I]{Lee_2021} & LLDUC orbitals \\  \hline
    19 & Simulating Nature & Chemistry & cytocrhome P450  & 1434 & & $7.8\times 10^{9}$ & \cite[Table IV]{Goings_2022} & $M=320$ \\ \hline 
    20 & Simulating Nature & Chemistry & EC molecule & 2685 & $6.32\times 10^{10}$ & & \cite[Table IV]{Kim_2022} & STO-3G basis \\ \hline 
    21 & Simulating Nature & Chemistry & EC molecule & 10462 & $5.41\times 10^{11}$ & & \cite[Table IV]{Kim_2022} & DZ basis \\ \hline 
    22 & Simulating Nature & Chemistry & EC molecule & 81958 & $6.25\times 10^{13}$ & & \cite[Table IV]{Kim_2022} & cc-pVTZ basis \\ \hline 
    23 & Simulating Nature & Materials & Ethylene Carbonate & 1395 & & $2.5\times 10^{10}$  & \cite[Table VII]{Su_2021} & plane wave basis \\ \hline 
    24 & Simulating Nature & Materials & LiPF$_{6}$ & 1758 & & $8\times 10^{10}$  & \cite[Table VII]{Su_2021} & ~"~ \\ \hline 
  25 & Simulating Nature & Materials & R$\bar{3}$M  & 166946 & & $6.16\times 10^{13}$  & \cite[Table VI]{rubin2023faulttolerant} & Sparse LCU \\ \hline 
  26 & Simulating Nature & Materials & C2/m  & 83532 & & $1.03\times 10^{13}$  & \cite[Table VI]{rubin2023faulttolerant} & ~"~ \\ \hline 
 27 & Simulating Nature & Materials & P2/c  & 99918 & & $2.06\times 10^{13}$  & \cite[Table VI]{rubin2023faulttolerant} & ~"~ \\ \hline 
28 & Simulating Nature & Materials & P$2_{1}$/c  & 182864 & & $3.39\times 10^{13}$  & \cite[Table VI]{rubin2023faulttolerant} & ~"~ \\ \hline 
29 & Simulating Nature & Materials & Fusion & 1749 & & $5.6\times 10^{14}$ & \cite[Table IV]{rubin2023quantum} & Line 1; QSP \\ \hline
 30 & Simulating Nature & Materials & Fusion & 2666 & & $1.1\times 10^{13}$ & \cite[Table IV]{rubin2023quantum} & Line 1; PF \\ \hline
 31 & Simulating Nature & Materials & Fusion & 33038 & & $2.1\times 10^{20}$ & \cite[Table IV]{rubin2023quantum} & Line 4; QSP \\ \hline
32 & Simulating Nature & Materials & Fusion & 33368 & & $2.1\times 10^{17}$ & \cite[Table IV]{rubin2023quantum} & Line 4; PF \\ \hline
33 & Simulating Nature & Chemistry & Chromium Dimer & 1366 & & $8\times 10^{9}$  & \cite[Fig II]{elfving2020will} & $N=26$ \\ \hline
34 & Simulating Nature & Chemistry & Ibrutinib & 2207 & & $1.1\times 10^{10}$  & \cite{Blunt_2022} & First sentence below Table 2. \\ \hline
35 & Simulating Nature & Molecular Forces & Water & 1790 & & $7.72\times10^{17}$  & \cite[Table II]{steudtner2023faulttolerant} & std-EVE \\ \hline
36 & Simulating Nature & Molecular Forces & Ammonia & 2130 & & $1.92\times10^{18}$  & \cite[Table II]{steudtner2023faulttolerant} & ~"~ \\ \hline
37 & Simulating Nature & Molecular Forces & p-Benzyne & 7577 & & $2.03\times10^{21}$  & \cite[Table II]{steudtner2023faulttolerant} & ~"~ \\ \hline
38 & Financial Engineering & PO & Asset Allocation & $10^{6}$ & $3.6\times 10^{7}$ & & \cite{egger2019credit} & $K=2^{20}$ on pages 4 \& 5 \\ \hline
39 & Financial Engineering & PO & Asset Allocation & $8\times 10^{6}$ & $2\times 10^{24}$ & & \cite{dalzell2023quantum} & $N=100$; interior point method \\ \hline
40 & Financial Engineering & RM & Derivative Pricing & 8000 & $1.2\times 10^{10}$ & & \cite[Table I]{chakrabarti2021threshold} & Auto-callable option; Re-parameterization method \\ \hline
41 & Financial Engineering & RM & Derivative Pricing & 11500 & $9.8\times 10^{9}$ & & \cite[Table I]{chakrabarti2021threshold} & TARF option; Re-parameterization method \\ \hline
42 & Financial Engineering & RM & Derivative Pricing & 4700 & $2.4\times 10^{9}$ & & \cite{stamatopoulos2023derivative} &  Last paragraph on page 11 \\ \hline
43 & Financial Engineering & RM & Derivative Pricing & 12000 & $1\times 10^{7}$ & & \cite{stamatopoulos2022towards} &  Last paragraph on page 17 \\ \hline
44 & Financial Engineering & RM & Option Pricing & $3.8\times 10^{4}$ & $1.9\times 10^{14}$ & & \cite{wang2023option} &  Table 6: Instance 1, Strong Euler scheme \\ \hline
45 & Financial Engineering & RM & Option Pricing & $2.2\times 10^{4}$ & $2.4\times 10^{11}$ & & \cite[Table 6]{wang2023option} & Instance 1, Weak Euler scheme \\ \hline
46 & Financial Engineering & RM & Option Pricing & $1.3\times 10^{5}$ & $7.7\times 10^{14}$ & & \cite[Table 6]{wang2023option} & Instance 4, Strong Euler scheme \\ \hline
47 & Financial Engineering & RM & Option Pricing & $6.4\times 10^{4}$ & $4.4\times 10^{11}$ & & \cite[Table 6]{wang2023option} & Instance 4, Weak Euler scheme \\ \hline
48 & Machine Learning & TDA & & 256 & & $6.8\times 10^{9}$ & \cite{berry2023analyzing} & Page 21 ($n=256, k=16$) \\ \hline
49 & Machine Learning & TDA & & 100 & & $1\times 10^{8}$ & \cite[Fig IIa]{berry2023analyzing} & $n=100, k=4$ \\ \hline
50 & Machine Learning & TDA & & 1000 & & $1\times 10^{10}$ & \cite[Fig IIa]{berry2023analyzing} & $n=1000, k=4$ \\ \hline
51 & Machine Learning & TDA & & 10000 & & $1\times 10^{12}$ & \cite[Fig IIa]{berry2023analyzing} & $n=10000, k=4$ \\ \hline
52 & Cryptography & ECC-110 & & 1014 & & $9.44\times 10^{9}$ & \cite[Table II]{roetteler2017quantum} & \\ \hline
53 & Cryptography & ECC-224 & & 2042 & & $8.43\times 10^{10}$ & \cite[Table II]{roetteler2017quantum} & \\ \hline 
54 & Cryptography & ECC-256 & & 2330 & & $1.26\times 10^{11}$ & \cite[Table II]{roetteler2017quantum} & \\ \hline
55 & Cryptography & ECC-521 & & 4719 & & $1.14\times 10^{12}$ & \cite[Table II]{roetteler2017quantum} & \\ \hline
56 & Cryptography & ECC-256 & & 2124 & & $7.4\times10^{9}$ & \cite[Table I]{haner2020} & Low W \\ \hline
57 & Cryptography & ECC-256 & & 2619 & & $2.3\times10^{9}$ & \cite[Table I]{haner2020} & Low $T$ \\ \hline
58 & Cryptography & ECC-384 & & 3151 & & $2.6\times10^{10}$ & \cite[Table I]{haner2020} & Low W \\ \hline
59 & Cryptography & ECC-384 & & 3901 & & $7.5\times10^{9}$ & \cite[Table I]{haner2020} & Low $T$ \\ \hline
60 & Cryptography & ECC-521 & & 4258 & & $6.2\times10^{10}$ & \cite[Table I]{haner2020} & Low W \\ \hline
61 & Cryptography & ECC-521 & & 5273 & & $1.7\times10^{10}$ & \cite[Table I]{haner2020} & Low $T$ \\ \hline
62 & Cryptography & ECC-256 & & 3000 & & $1.1\times 10^{8}$ & \cite[Figure II]{litinski2023compute} & \\ \hline
63 & Cryptography & ECC-256 & & 50000 & & $4.8\times10^{7}$ & \cite[Figure II]{litinski2023compute} & \\ \hline
64 & Cryptography & RSA-1024 & & 3078 &  $4\times 10^{8}$ & &  \cite[Table I]{gidney2021} & Cost is number of magic states \\ \hline
65 & Cryptography & RSA-2048 & & 6158 &  $2.7\times 10^{9}$  & & \cite[Table I]{gidney2021} &  " \\ \hline
66 & Cryptography & RSA-3072 & & 9237 &  $9.9\times 10^{10}$  & & \cite[Table I]{gidney2021} &  "\\ \hline

\caption{\textbf{Select Circuit-Model Resource Estimates.} The numbers above are taken from the references indicated. Note: CMP = Condensed Matter Physics, FH = Fermi-Hubbard Model, TFIM=Transverse Field Ising Model, SS=Spin Systems, HEP=High-Energy Physics, EFT = Effective Field Theory, QSP=Quantum Signal Processing, PF= Product Formula, PO=Portfolio Optimization, RM = Risk Management, LCU = Linear Combination of Unitaries, TDA=Topological Data Analysis}
    \label{tab:resource-estimates}
\end{longtable}

\newpage

\printbibliography
\end{document}